\documentclass[12pt]{article}
\usepackage{amsmath,amssymb}
\usepackage[english]{babel}
\usepackage[cp866]{inputenc}
\usepackage{hyperref}
\numberwithin{equation}{section}

\newcommand{\bq}{\begin{eqnarray}}
\newcommand{\eq}{\end{eqnarray}}
\newcommand{\bbq}{\begin{equation*}}
\newcommand{\eeq}{\end{equation*}}

\newcommand{\be}{\begin{equation}}
\newcommand{\ee}{\end{equation}}
\newcommand{\ra}{\rightarrow}

\newcommand{\la}{\Lambda_Q}
\newcommand{\mph}{\mu_{\Phi}}
\newcommand{\ov}{\overline}
\newcommand{\lym}{\Lambda_{YM}}

\newcommand{\wh}{\widehat}
\newcommand{\qq}{{\ov Q}Q}
\newcommand{\w}{{\cal W}_{\rm matter}}

\newcommand{\Qo}{({\ov Q}Q)_1}
\newcommand{\Qt}{({\ov Q}Q)_2}
\newcommand{\qo}{({\ov q}q)_1}
\newcommand{\qt}{{(\ov q}q)_2}

\newcommand{\mt}{\langle m_Q^{\rm tot}\rangle}

\newcommand{\bo}{{\rm b_o}}

\newcommand{\nd}{{\ov N}_c}
\newcommand{\mo}{\mu_{\Phi,\rm o}}

\newcommand{\mup}{\mu^{\rm pole}_q}

\newcommand{\mos}{\mu_o^{\rm str}}

\newcommand{\no}{{\rm n}_1}
\newcommand{\nt}{{\rm n}_2}

\newcommand{\mgo}{\mu^{\rm pole}_{gl,1}}
\newcommand{\mgt}{\mu^{\rm pole}_{gl,2}}

\begin{document}

\begin{center}{\bf \large Mass spectra in $\mathbf{{\cal N}=1\,\, SQCD}$ with additional colorless fields.\\
Strong coupling regimes. II } \end{center}
\vspace{1cm}
\begin{center}\bf Victor L. Chernyak $^{\,a,\,b}$ \end{center}
\begin{center}(e-mail: v.l.chernyak@inp.nsk.su) \end{center}
\begin{center} $^a\,$ Budker Institute of Nuclear Physics SB RAS \\
 630090 Novosibirsk, Russian Federation\end{center}

\begin{center} $^b\,$ Novosibirsk State University,\\ 630090 Novosibirsk, Russian Federation\end{center}
\vspace{1cm}
\begin{center}{\bf Abstract} \end{center}
\vspace{1cm}

This paper continues our studies in arXiV:1608.06452 [hep-th] of ${\cal N}=1$ gauge theories in the strongly coupled regimes. We also consider here in details the ${\cal N}=1$ SQCD-like theories with $SU(N_c)$ colors (and their Seiberg's dual), with  $N_F$ flavors of light quarks and $N_F^2$ additional colorless flavored scalars $\Phi^j_i$, but now with $N_F$ in the range $N_F>3N_c$.

The mass spectra of these direct and dual theories in various vacua are calculated within the dynamical scenario introduced by the author in \cite{ch3}. It assumes that quarks in such ${\cal N}=1$ SQCD-like theories without elementary colored adjoint scalars can be in two {\it standard} phases only. These are either the HQ (heavy quark) phase where they are confined or the Higgs phase. Recall that this scenario satisfies all those tests which were used as checks of the Seiberg hypothesis about the equivalence of the direct and dual theories. Calculated mass spectra of the direct $SU(N_c)$ theory are compared to those of its Seiberg's dual $SU(N_F-N_c)$ variant and appeared to be parametrically different.

\newpage

\tableofcontents

\newpage
\section{Introduction}

\hspace*{1cm} We continue in this article our previous study in \cite{ch7} of strongly coupled ${\cal N}=1$ SQCD-like theories with $SU(N_c)$ colors (and their Seiberg's dual with $SU(N_F-N_c)$ dual colors), with $N_F$ flavors of light quarks and with additional $N_F^2$ colorless but flavored fields $\Phi^j_i$. But this time considered is the region $N_F>3N_c$.

First of all, remind that the Lagrangian of the direct $SU(N_c)$ theory at the scale $\mu=\la$ has the form
\footnote{\,
The gluon exponents are always implied in the Kahler terms. Besides, here and everywhere below in the text we neglect for simplicity all RG-evolution effects if they are logarithmic only.
}
\bbq
K={\rm Tr}\, (\Phi^\dagger \Phi )+{\rm Tr}\Bigl (\,Q^\dagger Q+Q\ra{\ov Q}\,\Bigr )\,,\quad {\cal W}=-\frac{2\pi}{\alpha(\mu=\la)}S+{\cal W}_{\rm matter}\,,
\eeq
\bq
{\cal W}_{\rm matter}={\cal W}_{\Phi}+{\cal W}_Q\,,\quad{\cal W}_{\Phi}=\frac{\mph}{2}\Biggl [{\rm Tr}\,(\Phi^2)-\frac{1}{N_F-N_c}\Bigl ({\rm Tr}\,\Phi\Bigr )^2\Biggr ],\quad {\cal W}_Q={\rm Tr}\,{\ov Q}(m_Q-\Phi) Q\,.\label{(1.1)}
\eq

Here\,: $\mph\gg\la$ and $m_Q\ll\la$  are the mass parameters, the traces in \eqref{(1.1)} are over color and/or flavor indices, $S=-\sum_{A,\beta} W^{A}_{\beta}W^{A,\,\beta}/32\pi^2$, where $W^A_{\beta}$ is the gauge field strength, $A=1...N_c^2-1,\, \beta=1,2$,\, $a(\mu)=N_c \alpha(\mu)/2\pi=N_c g^2(\mu)/8\pi^2$ is the running gauge coupling with its scale factor $\la$, $\,Q^i_a, {\ov Q}_j^{\,a},\,\,a=1...N_c,\,\,
i,j=1...N_F$ are the quark fields, $\Phi^j_i$ are $N_F^2$ colorless flavored scalars. {\it This normalization of fields is used everywhere below in the main text}. Besides, the perturbative NSVZ $\beta_{NSVZ}$-functions for (effectively) massless SUSY theories \cite{NSVZ1,NSVZ2} are used in this paper.

There is a large number of various type vacua in this theory and for a reader convenience we reproduce in Appendix the results from section 2 in \cite{ch6} for the mean vacuum values ("condensates")\, of quarks and gluino,  $\langle{\ov Q}_j Q^i\rangle\equiv\sum_{a=1}^{N_c}\langle{\ov Q}^{\,a}_j Q_a^i\rangle$
and $\langle S\rangle$, in all different vacua at $N_F>2N_c\,$.

This direct $SU(N_c)$ theory with $N_F>3N_c$ quark flavors and $m_Q\ll\la$ is IR free and logarithmically weakly coupled at scales $\mu<\la$, $\,a(\mu\ll\la)\sim 1/\log (\la/\mu)\ll 1$, while it is strongly coupled at $\mu>\la,\,\, a(\mu\gg\la)\sim (\la/\mu)^{\nu_Q\,<\,0}\gg 1$.

In parallel with this direct $\Phi$-theory \eqref{(1.1)}, we study also its Seiberg's dual variant \cite{S2,IS}, the $d\Phi$-theory with $SU(\nd=N_F-N_c)$ dual colors, $N_F$ flavors of dual quarks $q_i^b, {\ov q}^{\,j}_b,\,\,b=1...\nd$, and with $N_F^2$ additional colorless but flavored elementary fields 
$M^i_j\ra ({\ov Q}_j Q^i)$. Its Lagrangian at $\mu=\la$ looks as, see \eqref{(1.1)} for ${\cal W}_{\Phi}$,
\bbq
{\ov K}={\rm Tr}\,(\Phi^\dagger\Phi)+ {\rm Tr}\,\Bigl (\frac{M^{\dagger}M}{\la^2}\Bigr )+{\rm Tr}\,\Bigl ( q^\dagger q + q\ra{\ov q}\Bigr )\,,\quad
\ov{\cal W}=\, -\,\frac{2\pi}{{\ov\alpha}(\mu=\la)}\,{\ov S}+{\ov{\cal W}}_{\rm matter}\,,
\eeq
\bq
{\ov{\cal W}}_{\rm matter}={\cal W}_{\Phi}+{\cal W}_{\Phi M}+{\cal W}_q\,,\quad {\cal W}_{\Phi M}={\rm Tr}\,(m^{\rm tot}_Q M)=
{\rm Tr}\,(m_Q-\Phi)M\,,\quad {\cal W}_q= -\,\rm {Tr}\, \Bigl ({\ov q}\,\frac{M}{\la}\, q \Bigr )\,. \label{(1.2)}
\eq
Here\,:\, the number of dual colors is ${\ov N}_c=(N_F-N_c),\,\,{\ov q}^{j}_b, q^{b}_i$ are dual quark fields, $b=1...N_F-N_c,\,\, M^i_j$ are $N_F^2$ Seiberg's elementary mion fields, $M^i_j\ra ({\ov Q}_j Q^i)$,\,\, ${\ov a}(\mu)=\nd{\ov \alpha}(\mu)/2\pi=\nd{\ov g}^2(\mu)/8\pi^2$ is the dual running gauge coupling (with its scale parameter $\Lambda_q$),\,\,${\ov S}=-\sum_{B,\beta}{\rm
\ov W}^{\,B}_{\beta}\,{\rm \ov W}^{\,B,\,\beta}/32\pi^2$,\,\, ${\rm \ov W}^{\,B}_{\beta}$ is the dual gluon field strength, $B=1...\nd^2-1$. The gluino condensates of the direct and dual theories are matched, $\langle{-\,\ov S}\rangle=\langle S\rangle\equiv\lym^3$, as well as $\langle M^i_j\rangle\equiv\langle M^i_j(\mu=\la)\rangle=\langle{\ov Q}_j Q^i (\mu=\la)\rangle\equiv\langle{\ov Q}_j Q^i\rangle$, and the scale parameter $\Lambda_q$ of the dual gauge coupling is taken as $|\Lambda_q|=\la$.

In contrast to the direct $SU(N_c)$ theory \eqref{(1.1)}, this dual $SU(\nd=N_F-N_c)$ theory with $N_F>3N_c$ is UV free  and logarithmically weakly coupled at $\mu>\la,\,\, {\ov a}(\mu\gg\la)\sim 1/\log(\mu/\la)\ll 1$. But it is strongly coupled at $\mu\ll\la$, with ${\ov a}(\mu\ll\la)\sim (\la/\mu)^{\nu_q\,>\,0}\gg 1$.

If all $N_F^2$ fields $\Phi^j_i$ are too heavy and dynamically irrelevant, they all can be then integrated out and, instead \eqref{(1.2)}, the superpotential will be
\bq
{\ov{\cal W}}_{\rm matter}={\cal W}_{M}+{\cal W}_q\,,\quad {\cal W}_{M}=m_Q\,{\rm Tr\,} M-\frac{1}{2\mph}\Biggl ({\rm Tr\,} (M^2)-\frac{1}{N_c}({\rm Tr\,} M)^2\, \Biggr )\,.\label{(1.3)}
\eq

As in the previous article \cite{ch7}, our purposes here is\,: a) to calculate mass spectra in both direct and dual theories \eqref{(1.1)},\eqref{(1.2)},\, b) to compare these two mass spectra in order to see whether they are the same or not.

Recall that in the region $N_c<N_F<3N_c/2,\, m_Q\ll\la,\, \mph\ll\la^2/m_Q$ considered in \cite{ch7}, all quark and gluon masses in {\it all vacua} of both theories \eqref{(1.1)},\eqref{(1.2)} were much less than $\la$, so that in all vacua we have dealt with the direct theory \eqref{(1.1)} in the strong coupling regime with $a(\mu\ll\la)\gg 1$, while the dual theory was in the weak coupling one, with ${\ov a}(\mu\ll\la)\ll 1$. As will be seen from the text below, this is not the case for the region $N_F>3N_c$ considered in this paper. It turns out that\,: a) in some vacua the quark and gluon masses are much larger than $\la$, so that we have to deal with the strongly coupled direct theory \eqref{(1.1)} and the weakly coupled dual one \eqref{(1.2)},\, b) while in other vacua the quark and gluon masses are much smaller than $\la$, so that we have to deal with the weakly coupled direct theory and strongly coupled dual one.\\

Recall also that, as was argued in detail in section 7 of \cite{ch1}, outside the conformal window $3N_c/2<N_F<3N_c$, the standard UV free direct theory \eqref{(1.1)} (i.e. without fields $\Phi$) with $N_c<N_F<3N_c/2$ and $m_Q\ll\la$ enters smoothly at scales $\mu<\la$ into {\it the strongly coupled effectively massless perturbative regime} with the gauge coupling $a(\mu\ll\la)\sim (\la/\mu)^{\,\nu_Q>\,0}\gg 1$. In short, the arguments were as follows.\\
{\bf a)} It was proposed by Seiberg in \cite{S2} that the standard direct (electric) ${\cal N}=1\,\, SU(N_c)$ SQCD (i.e. \eqref{(1.1)} without fields $\Phi$)
at $N_c<N_F<3N_c/2$ and with $m_Q= 0$ is in the "confinement without spontaneous chiral symmetry breaking" (and without R-symmetry breaking) regime (\, $\langle{\ov Q}_j Q^i\rangle=\langle S\rangle=\langle{\ov q}^{\,j} q_i\rangle=\langle M^i_j\rangle=0\,$). And the standard massless dual (magnetic) $SU(N_F-N_c)$ theory \eqref{(1.2)} (i.e. without fields $\Phi$) was proposed as its low energy form at scales $\mu<\la$. This implies: a) the confinement of all original electric quarks and gluons with the string tension $\sqrt{\sigma}\sim\la$ (with $\la$ the only dimensional parameter), b) all colorless hadrons made from electric quarks and gluons have masses $\sim\la$, c) the formation of solitonic IR free $SU(N_F-N_c)$ theory with massless at $m_Q=0$ magnetic quarks and gluons (and $M^i_j$ fields). Besides, both direct and dual theories are considered in \cite{S2} as nonsingular nontrivial interacting theories at $m_Q=0$.\\
{\bf b)} As was argued in section 7 of \cite{ch1}, the problem with this scenario is that it is impossible to write at the scale $\mu\sim\la$ the effective Lagrangian of massive hadrons with masses $\sim\la$ (made of massless confined electric quarks and gluons), which will be nonsingular at $m_Q=0$ and preserving both the R-symmetry and chiral symmetry $SU(N_F)_L\times SU(N_F)_R\,$. The variant with possibly massive (confined or not) electric quarks with masses $\sim\la$ has the same problems as those for hadrons. The nonsingular dynamically induced quark mass term in the superpotential at the scale $\mu=\la$ will look then as $\delta{\cal W}\sim\sum_{i,j=1}^{N_F}\langle\phi^j_i\rangle\,{\ov Q}_j Q^i,\,\,\langle\phi^j_i\rangle\sim\delta_i^j\la$, with some composite field $\phi^j_i$. But, in any case, this will break spontaneously the exact chiral symmetry $SU(N_F)_L\times SU(N_F)_R$ of the $m_Q=0$ theory. Besides, this (solitonic colorless or colorful) field $\phi$ has the positive R-charge, $R_{\phi}=2N_c/N_F$. But, first, $\langle\phi^j_i\rangle\sim\delta_i^j\la$ will break spontaneously also the R-symmetry of the $m_Q\ll\la$ theory, while R-symmetry is considered in \cite{S2} as unbroken in both direct and dual theories (there is no corresponding massless Nambu-Goldstone boson). And second, the unbroken at $m_Q\ll\la$ R-symmetry requires that any chiral (elementary or composite) superfield $\Psi$ with the R-charge $r > 0\,$ behaves as $\langle\Psi(\mu=\la)\rangle\sim (m_Q)^{r N_F/2N_c}\ra 0$ at $m_Q\ra 0$, and this forbids $\langle\phi^j_i\rangle\sim\delta_i^j\la\,$.

We also recall here the following. There is no confinement in Yukawa-like theories without gauge interactions. The confinement originates {\it only} from the unbroken YM, or ${\cal N}=1$ SYM in ${\cal N}=1$ SQCD-like theories. And because ${\cal N}=1$ SYM has only one dimensional parameter $\langle\lym\rangle=\langle S\rangle^{1/3}$, the string tension is $\sigma^{1/2}\sim\langle\lym\rangle$. But in the standard ${\cal N}=1$ SQCD with light quarks the value of $\lym$ is well known: $\lym=(\Lambda_Q^\bo\det m_Q)^{1/3N_c}\ll\Lambda_Q$. Therefore, in contradiction with Seiberg's assumption in \cite{S2} about confinement with the string tension $\sigma^{1/2}\sim\la\gg\lym$, the ${\cal N}=1$ SYM cannot produce confinement with the string tension $\sim\Lambda_Q$ (and there is no confinement at all at $m_Q\ra 0$).

Besides, within the mentioned above dynamical scenario introduced in \cite {ch3}, it was shown in section 7 of \cite{ch3} that at $N_c<N_F<3N_c/2$ the mass spectra of the direct and dual theories are qualitatively different. In this scenario \cite{ch3}, according to reasonings given above, the direct standard 
${\cal N}=1$ SQCD with light quarks, i.e. $m_Q\ll\la$, enters smoothly at $\mu<\la$ into the strongly coupled effectively massless perturbative regime with $a(\mu\ll\la)\sim (\la/\mu)^{\,\nu_Q>\,0}\gg 1$. The IR free massless Seiberg's dual theory is considered as {\it independent theory} which presumably becomes equivalent to the direct one at $\mu<\la$. As was shown in \cite{ch3}, the mass spectra of the direct and dual theories are not equivalent in this case.\\

As was argued in section 7 of \cite{ch1} (see also Appendix in \cite{ch6} for the anomalous dimension $\gamma_{\Phi}$ of the field $\Phi$ in the direct theory), the anomalous dimensions $\gamma_Q$ of the quark field $Q$ and $\gamma_q$ of the dual quark field $q$  are related by
\bq
\nd (1+\gamma_Q)=N_c (1+\gamma_q)\,. \label{(1.4)}
\eq

From \eqref{(1.4)} at $N_F>3N_c$\,. -\\
a) At $\mu\gg\la$ where the dual UV free theory is weakly coupled with $\gamma_q \ra 0$ while the IR free direct theory is strongly coupled:
\bq
\gamma_Q=\frac{2N_c-N_F}{\nd}\,\,,\quad \gamma_{\Phi}=\,-\,2\gamma_Q\,,\quad \nd=N_F-N_c\,,\label{(1.5)}
\eq
\bbq
\frac{d\,a(\mu)}{d\log\mu}=\beta_{NSVZ}(a)=\frac{a^2(\mu)}{a(\mu)-1}\,\frac{\bo-N_F\gamma_Q}{N_c}\quad
\xrightarrow{a(\mu)\gg 1}\quad -\,\nu_Q\,a(\mu),\quad \bo=3N_c-N_F\,,
\eeq
\bbq
\nu_Q=\frac{N_F\gamma_Q-\bo}{N_c}=\frac{3N_c-2N_F}{\nd}=\gamma_Q-1\,<\,0\,,\quad
a(\mu\gg \la)\sim\Bigl (\frac{\la}{\mu}\Bigr )^{\nu_Q}\gg 1\,.
\eeq
From \eqref{(1.5)}, in the region $N_F>3N_c$ considered in this paper and at $\mu>\la$ \,: $\, -1<\gamma_Q<0\,,\,\,0<\gamma_{\Phi}<2\,,\,\, -2<\nu_Q<\,-1\,$. Besides, the gauge coupling of the $SU(N_c)$ SYM looks in the strong coupling regime as \cite{ch1}
\bbq
\frac{d\, a^{\rm str}_{YM}(\mu\gg\lym)}{d\log\mu}=\beta^{YM}_{NSVZ}(a^{\rm str}_{YM})=\frac{3( a^{\rm str}_{YM})^2}{a^{\rm str}_{YM}-1}\,\,\,\,
\xrightarrow{a^{\rm str}_{YM}\gg 1}\,\,\, 3\, a^{\rm str}_{YM}(\mu)\,,
\eeq
\bq
a^{\rm str}_{YM}(\mu\gg\lym)\sim\Bigl (\frac{\mu}{\lym}\Bigr )^3\gg 1\,. \label{(1.6)}
\eq
b) For the strongly coupled at $\mu\ll\la$ dual $SU(\nd=N_F-N_c)$ theory \eqref{(1.2)} with $N_F>3N_c$ the corresponding anomalous dimensions $\gamma_q\,,\,\gamma_M\,,\,\gamma_{\Phi}$ and the dual gauge coupling ${\ov a}(\mu)$ look as
\bq
\gamma_q=\frac{2\nd-N_F}{N_F-\nd}=\frac{N_F-2N_c}{N_c}\,>\,1\,,\quad \gamma_M=\,-2\gamma_q\,,\quad \nu_q=\gamma_q-1=\frac{N_F-3N_c}{N_c}\,>\,0\,,\quad \gamma_{\Phi}=0\,,\label{(1.7)}
\eq
\bq
{\ov a}(\mu\ll\la)\sim\Bigl (\frac{\la}{\mu}\Bigr )^{\,\nu_q}\gg 1\,,
\quad {\ov a}^{\,\rm str}_{YM}(\mu\gg\lym)\sim\Bigl (\frac{\mu}{\lym}\Bigr )^3\gg 1\,.\label{(1.8)}
\eq

Recall also that the parametric behaviour and hierarchies of condensates are qualitatively different at $\mph\lessgtr\mo=\la(\la/m_Q)^{(2N_c-N_F)/N_c}$, see section 3 in \cite{ch4} and section 2 in \cite{ch6}. Therefore, because we consider in this paper the region $N_F>3N_c$ and $\mph\gg\la,\, m_Q\ll\la$, then $\mo\ll\la$ is very small and we always deal with the region $\mph\gg\mo$ in this paper.\\

For calculations of mass spectra in various vacua in the strongly coupled regimes we use the dynamical scenario introduced in \cite{ch3}. This scenario assumes that quarks in such ${\cal N}=1$ SQCD-like theories can be in two {\it standard} phases only\,: these are either the HQ (heavy quark) phase where they are not higgsed but confined, with $\langle Q\rangle=\langle{\ov Q}\rangle=0$, or the Higgs phase where they form nonzero coherent condensate with (at least some components of) $\langle Q^i_a\rangle=\langle{\ov Q}_i^a\rangle\neq 0$, breaking the color symmetry. The word {\it standard} implies here also that, unlike ${\cal N}=2$ SQCD, in such ${\cal N}=1$ theories without elementary colored adjoint scalars, no additional cparametrically lighter solitons (e.g. magnetic monopoles or dyons) are formed at those scales where quarks decouple as heavy or are higgsed.
\footnote{\,
In the logarithmically weakly coupled regimes the dynamics is simple and clear and we need no additional assumptions because this scenario is literally standard.
\label{(f2)}
}
\\

The organization of the paper is as follows. The vacua with the unbroken $U(N_F)$ global flavor symmetry are considered in sections 2 and 3, in both the direct and dual theories. The vacua with the spontaneously broken $U(N_F)_{L+R}\ra U(\no)_{L+R}\times U(\nt)_{L+R}$ global flavor symmetry are considered in sections 4-6. In Appendix, for a reader convenience, we remind from \cite{ch6} the values of various condensates in different vacua at $N_F>2N_c$.
\vspace*{4mm}

\addcontentsline{toc}{section}
 {\hspace*{3cm} Unbroken flavor symmetry $U(N_F)$}
{\bf \Large\hspace*{2cm} Unbroken flavor symmetry $U(N_F)$}

\section{QCD vacua}

There are $N_c$ quantum QCD vacua at $\mph\gg\la\gg\mo=\la (m_Q/\la)^{(N_F-2N_c)/N_c}$. The condensates in the direct theory look in these vacua as, see Appendix,
\bbq
\langle\qq\rangle_{QCD}=\langle\qq(\mu=\la)\rangle_{QCD}\simeq\frac{\langle S\rangle_{\rm QCD}\equiv \langle\lym^{(\rm QCD)}\rangle^3}{m_Q}\simeq\frac{1}{m_Q}\Bigl (\la^{\bo}m_Q^{N_F}\Bigr)^{1/N_c}\,,\quad \frac{\langle\qq\rangle_{QCD}}{\la^2}\ll 1\,,
\eeq
\bq
\mt=\langle m_Q-\Phi\rangle=\frac{\langle S\rangle_{\rm QCD}}{\langle\qq\rangle_{QCD}}\simeq m_Q\ll\la\,,\quad \langle\Phi\rangle\sim\frac{\langle\qq\rangle_{QCD}}{\mph}\ll m_Q\,.\label{(2.1)}
\eq

The condensates in the dual theory look as
\bq
\langle M\rangle_{QCD}\equiv\langle M(\mu=\la)\rangle_{QCD}=\langle\qq\rangle_{QCD},\quad \langle {\ov q} q\rangle_{QCD}=\frac{\la\langle S\rangle_{\rm QCD}}{\langle M\rangle_{QCD}}\simeq m_Q\la\ll\la^2\,.\label{(2.2)}
\eq

\subsection{Direct theory}

All quark and gluon masses are $\ll\la$ in this IR free logarithmically weakly coupled at $\mu<\la$ direct $SU(N_c)$ theory, see \eqref{(2.1)}. But something nontrivial takes place with masses of $N_F^2$ fions $\Phi^j_i$. This direct theory with $N_F>3N_c$ is strongly coupled at scales $\mu\gg\la$, with the gauge coupling $a(\mu\gg\la)\sim (\la/\mu)^{\nu_Q\,<\,0}\gg 1$, see \eqref{(1.4)}. The running mass $\mu_{\Phi}(\mu)$ of $\Phi$ looks as, see \eqref{(1.4)},
\bq
\mu_{\Phi}(\mu=\la)=\mph\gg\la\,,\,\, \mu_{\Phi}(\mu\gg\la)=\frac{\mph}{z_{\Phi}(\la,\mu)}\,,\,\,z_{\Phi}
(\la,\mu)=\Bigl (\frac{\mu}{\la}\Bigr )^{\gamma_{\Phi}}\gg 1\,,\,\, \gamma_{\Phi}=-2\gamma_Q\,>\,1\,. \label{(2.3)}
\eq
Therefore, the running mass $\mu_{\Phi}(\mu)$ of fions {\it decreases} with increasing $\mu>\la$ and (if nothing prevents) $\mu_{\Phi}(\mu>\mos)<\mu$, so that all $N_F^2$ fion fields $\Phi^j_i$ {\it become relevant} at scales $\mu>\mos$, see section 4 in \cite{ch4} and Appendix in \cite{ch6},
\bq
\la\ll\mu_{o}^{str}=\la\Bigl (\frac{\la}{\mph}\Bigr )^{\frac{1}{2\gamma_Q-1}}=\la\Bigl (\frac{\mph}{\la}
\Bigr )^{\frac{\nd}{3N_F-5N_c}}\ll\mph\,. \label{(2.4)}
\eq
Thus, when we are going up in energy from $\mu=\la$, the fions are dynamically  irrelevant in the interval of scales $\la<\mu<\mos\ll\mph$ where $\mu_{\Phi}(\mu)>\mu$, but {\it become relevant} at $\mu>\mos$ where $\mu_{\Phi}(\mu)<\mu\,$. This means that, because $\mu_{\Phi}(\mu=\mph)\ll\mph$, there is no pole in propagators of fions at $\mu\sim\mph$, but there is a pole at $\la\ll\mu^{\rm pole}(\Phi)\sim\mos\ll\mph$.

At $\mu<\mos$ all fions become too heavy and dynamically irrelevant. Therefore, they all can be integrated out at $\mu<\mos$, see \eqref{(2.4)}, and the superpotential takes the form
\bq
{\cal W}_Q=m_Q{\rm Tr}({\ov Q} Q)-\frac{1}{2\mph}\Biggl (\,\sum_{i,j}\,({\ov Q}_j Q^i)({\ov Q}_i Q^j)-\frac{1}{N_c}\Bigl({\rm Tr}\,{\ov Q} Q \Bigr)^2 \Biggr ). \label{(2.5)}
\eq
The last term in \eqref{(2.5)} is irrelevant at scales $\mu<\la$ (up to small power corrections). Therefore, the mass spectrum of the direct theory at $\mu<\la$ will be the same as in the standard ${\cal N}=1$ SQCD without fields $\Phi$. The quark mass is (up to ignored parametrically large logarithmic factor, see the footnote 1) $m_Q^{\rm pole}\sim m_Q$. The gluon mass due to possible higgsing of quarks is: $\mu_{gl}\sim
\langle\qq\rangle^{1/2}_{QCD}\ll m_Q$, see \eqref{(2.1)}. Therefore, {\bf the overall phase is HQ} (heavy quarks). All quarks decouple as heavy in the weak coupling region at $\mu<m_Q^{\rm pole}\ll\la$. There remains $SU(N_c)$ SYM. The scale factor $\lambda_{YM}$ of its gauge coupling is determined from matching $a_{(+)}(\mu=m_Q^{\rm pole})=a_{YM}(\mu=m_Q^{\rm pole})$ and is $\lambda_{YM}=\langle\lym^{(\rm QCD)}\rangle$ as it should be, see \eqref{(2.1)}.\\

On the whole, the spectrum of masses looks as follows. -

1) The quark mass is $m_Q^{\,\rm pole}\sim m_Q$ (up to logarithmic factor). All quarks are weakly coupled and weakly confined (i.e. the tension of the confining string originating from $SU(N_c)$ SYM is $\sqrt{\sigma}\sim\langle\lym^{(\rm QCD)}\rangle\ll m_Q$), see \eqref{(2.1)}. Therefore, there is a large number of hadrons made from these weakly coupled non-relativistic quarks, their mass scale is $\mu_{\rm H}\sim m_Q\ll\la$.\\
2) There is a large number of strongly coupled gluonia with the mass scale $\sim\langle\lym^{(\rm QCD)}\rangle\ll m_Q$, see \eqref{(2.1)}.\\
3) All $N_F^2$ fions $\Phi^j_i$ have large masses $\la\ll\mu^{\rm pole}(\Phi)\sim\mos$, see \eqref{(2.4)}.

\subsection{Dual Theory}

The dual theory is strongly coupled at $\mu<\la$, with the dual gauge coupling ${\ov a}(\mu=\la)\sim 1,\,\,
{\ov a}(\mu\ll\la)\sim (\la/\mu)^{\nu_q>\,0}\gg 1$. The dual Lagrangian at scales $\mup<\mu<\la$ has the form, see \eqref{(1.2)},\eqref{(1.6)},
\bq
K={\rm Tr\,}\Bigl [ \Phi^\dagger\Phi+z_M(\la,\mu)\frac{M^\dagger M}{\la^2}+z_q(\la,\mu)\Bigl (q^\dagger q+{\ov q}^\dagger {\ov q}\Bigr ) \Bigr ]\,,
\label{(2.6)}
\eq
\bbq
\w={\cal W}_{\Phi}+{\rm Tr\,}\Bigl [m^{\rm tot}_Q M-{\ov q}\,\frac{M}{\la}q  \Bigr ],\quad {\cal W}_{\Phi}
=\frac{\mph}{2}\Bigl [{\rm Tr\,}(\Phi^2)-\frac{1}{\nd}\Bigl ({\rm Tr\,}\Phi \Bigr )^2 \,\Bigr ], \quad m^{\rm tot}_Q=m_Q-\Phi\,,
\eeq
\bbq
z_q(\la,\mu)=\Bigl (\frac{\mu}{\la}\Bigr )^{\gamma_q}\,,\quad z_M(\la,\mu)=\Bigl (\frac{\mu}{\la}\Bigr )^{\gamma_M=-2\gamma_q}\,,\quad \gamma_q=\frac{2\nd-N_F}{N_F-\nd}=\frac{N_F-2N_c}{N_c}>\,0,\quad \gamma_{\Phi}=0\,.
\eeq

All $N_F^2$ fields $\Phi$ are heavy, $\mu^{\rm pole}(\Phi)\sim\mph\gg\la$. Therefore, they all are  dynamically irrelevant at $\mu<\mph$ and can be integrated out. The dual Lagrangian becomes,
\bq
K={\rm Tr\,}\Bigl [z_M(\la,\mu)\frac{M^\dagger M}{\la^2}+z_q(\la,\mu)\Bigl (q^\dagger q+{\ov q}^\dagger {\ov q}\Bigr ) \,\Bigr ]\,,\label{(2.7)}
\eq
\bbq
\w={\cal W}_{M}-{\rm Tr\,}\Bigl (\,{\ov q}\,\frac{M}{\la}q\,\Bigr )\,,\quad {\cal W}_{M}=m_Q{\,\rm Tr\,}(M)
-\frac{1}{2\mph}\Bigl [{\rm Tr\,}(M^2)-\frac{1}{N_c}\Bigl ({\rm Tr\,} M \Bigr )^2 \,\Bigr ]\,.
\eeq

The potentially important masses look here as follows. The pole mass of dual quarks, see \eqref{(1.6)},
\bq
\mu^{\rm pole}_q\sim\frac{1}{z_q(\la,\mu=\mu^{\rm pole}_q)}\frac{\langle M\rangle}{\la}\quad\ra\quad \mu^{\rm pole}_q\sim m_Q\,.\label{(2.8)}
\eq
The gluon mass due to possible higgsing of dual quarks, see \eqref{(1.6)},\eqref{(1.7)},
\bbq
\Bigl ({\ov\mu}^{\,\rm pole}_{gl}\Bigr )^2\sim\Bigl [{\ov a}(\mu={\ov\mu}^{\,\rm pole}_{gl})=\Bigl (\frac{\la}{{\ov\mu}^{\,\rm pole}_{gl}} \Bigr )^{\nu_q}\Bigr ]z_q(\la,\mu={\ov\mu}^{\,\rm pole}_{gl})\,\langle{\ov q} q\rangle\,,\quad {\ov a}({\ov\mu}^{\,\rm pole}_{gl})z_q(\la,{\ov\mu}^{\,\rm pole}_{gl})\sim\frac{{\ov\mu}^{\,\rm pole}_{gl}}{\la}\,,
\eeq
\bq
{\ov\mu}^{\,\rm pole}_{gl}\sim\frac{\langle{\ov q} q\rangle}{\la}\sim m_Q \sim \mu^{\rm pole}_q\,,
\quad \quad \nu_q=\frac{3\nd-2N_F}{N_F-\nd}=\frac{N_F-3N_c}{N_c}\,>\,0\,. \label{(2.9)}
\eq

Because the global flavor symmetry $U(N_F)$ is unbroken in these vacua, this shows that dual quarks are not higgsed. Otherwise, due to rank restriction at 
$N_F > 3 N_c$, the global $U(N_F)$ would be broken spontaneously. Therefore, {\bf the overall phase is Hq} (heavy quarks). After integrating out all dual quarks as heavy at $\mu<\mu^{\rm pole}_q$ there remains $SU(\nd)$ SYM in the strong coupling regime. The scale factor of its gauge coupling is determined from the matching, see \eqref{(1.7)},\eqref{(2.8)},
\bbq
{\ov a}_{\,(+)}(\mu=\mu^{\rm pole}_q)=\Biggl (\frac{\la}{\mu^{\rm pole}_q}\Biggr )^{\nu_q}={\ov a}_{\,YM}(\mu=\mu^{\rm pole}_q)=\Bigl (\frac{\mu^{\rm pole}_q}{\lambda_{YM}}\Bigr )^3\quad
\ra
\eeq
\bq
\quad\ra\quad\lambda_{YM}=\Biggl (\la^{3\nd-N_F}\langle\det\frac{M}{\la}\rangle\Biggr )^{1/3\nd}=\langle\lym^{(\rm QCD)}\rangle\,, \label{(2.10)}
\eq
as it should be, see \eqref{(2.1)}. After integrating out dual $SU(\nd)$ gluons at $\mu<\langle\lym^{(\rm QCD)}\rangle$ via the VY procedure \cite{VY,TVY}, the Lagrangian looks as
\bq
K={\rm Tr}\Bigl [\, z_M(\la,\mu^{\rm pole}_q)\,\frac{M^\dagger M}{\la^2}\,\Bigr ]\,,\quad\quad
\w={\cal W}_{M}+{\cal W}_{\rm non-pert}\,, \label{(2.11)}
\eq
\bbq
{\cal W}_{\rm non-pert}=-\,\nd\Biggl (\la^{3\nd-N_F}\det\frac{M}{\la}\Biggr )^{1/\nd}\,.
\eeq
From \eqref{(2.11)}
\bq
\mu^{\,\rm pole}(M)\sim\frac{1}{z_M(\la,\mu=\mu^{\rm pole}_q)}\,\frac{\la^2\langle S\rangle}{\langle M\rangle^2}\sim\la\Bigl (\frac{m_Q}{\la}\Bigr )^{\frac{N_F-2N_c}{N_c}}\,,\quad
\frac{\mu^{\,\rm pole}(M)}{\langle\lym^{(\rm QCD)}\rangle}\sim\Bigl (\frac{m_Q}{\la}\Bigr )^{\frac{2(N_F-3N_c)}{3N_c}}\ll 1 \label{(2.12)}
\eq
(the main contribution to $\mu^{\,\rm pole}(M)$ originates from ${\cal W}_{\rm non-pert}$ in \eqref{(2.11)}.\\

On the whole, the mass spectrum of the dual theory look in these QCD-vacua as follows. -

1) The mass of dual quarks is $\mu^{\rm pole}_q\sim m_Q$, i.e. $\mu^{\rm pole}_q\sim m^{\rm pole}_Q$ (but only up to parametric logarithmic factor), see \eqref{(2.8)} and section 2.1. All quarks are weakly confined, i.e. the tension of the confining string originating from $SU(\nd)$ SYM is $\sqrt{\sigma}\sim\langle\lym^{(\rm QCD)}\rangle\ll m_Q$.\\
2) There is a large number of strongly coupled gluonia, their mass scale is $\sim\langle\lym^{(\rm QCD)}\rangle$.\\
3) Masses of $N_F^2$ mions $M^i_j$ are $\mu^{\,\rm pole}(M)\sim\la (m_Q/\la)^{(N_F-2N_c)/N_c}\ll\langle
\lym^{(\rm QCD)}\rangle$, see \eqref{(2.12)}.\\
4) There are $N_F^2$ fions $\Phi^j_i$ with masses $\mu^{\rm pole}(\Phi)\sim\mph$.

Comparing with the direct theory in section 2.1 it is seen that the mass spectra are parametrically different.

\section{L vacua}

There are $(N_F-2N_c)$ quantum L - vacua at $\mph\gg\mo$ with condensates, see Appendix,
\bq
\langle\qq\rangle_L\equiv\langle\qq(\mu=\la)\rangle_L\sim \la^2\Biggl (\frac{\mph}{\la}\Biggr )^{\frac{\nd}{N_F-2N_c}}\gg\la^2\,, \label{(3.1)}
\eq
\bbq
\langle\lym^{(L)}\rangle^3\equiv\langle S\rangle_L=\Bigl (\frac{\det\langle\qq\rangle_L}{\la^{3N_c-N_F}}\Bigr )^{1/\nd}\sim\la^3
\Bigl (\frac{\mph}{\la}\Bigr )^{\frac{N_F}{N_F-2N_c}}\gg\la^3\,.
\eeq
\bbq
\langle m_Q^{\rm tot}=m_Q-\Phi\rangle_L=\frac{\langle S\rangle_L}{\langle\qq\rangle_L}\sim\la
\Bigl (\frac{\mph}{\la}\Bigr )^{\frac{N_c}{N_F-2N_c}}\gg\la\,,
\eeq
\bbq
\langle M\rangle_L\equiv\langle M(\mu=\la)\rangle_L=\langle\qq\rangle_L\,,\quad\langle{\ov q} q\rangle=\frac{\la\langle S\rangle_L}{\langle\qq\rangle_L}
\sim\la^2\,.
\eeq

\subsection{Direct theory}

Because both $\langle m_Q^{\rm tot}\rangle_L\gg\la$ and $\langle\qq\rangle_L\gg\la^2$, we have to deal in this case with the direct theory in the strong coupling regime $a(\mu\gg\la)\gg 1$. The potentially important masses look here as follows. The quark pole mass, see \eqref{(1.4)},\eqref{(3.1)},
\bq
m_Q^{\,\rm pole}\sim\frac{\langle m_Q^{\rm tot}\rangle_L}{z_Q(\la,m_Q^{\,\rm pole})}\sim \la\Biggl (\frac{\mph}{\la}\Biggr )^{\frac{\nd}{N_F-2N_c}}\gg\mph\gg\la\,,\label{(3.2)}
\eq
\bq
z_Q(\la,m_Q^{\,\rm pole})\sim\Biggl (\frac{m_Q^{\,\rm pole}}{\la}\Biggr )^{\gamma_Q}\,, \quad -1\,<\gamma_Q=\frac{2N_c-N_F}{\nd}\,<\,-\,\frac{1}{2}\,.\quad \label{(3.3)}
\eq
The gluon mass due to possible higgsing of quarks, see \eqref{(1.4)},
\bbq
\Bigl (\mu^{\,\rm pole}_{gl}\Bigr )^2\sim\Biggl [a(\mu=\mu^{\,\rm pole}_{gl})\sim\Bigl (\frac{\la}{\mu^{\,\rm pole}_{gl}}\Bigr )^{\nu_Q} \Biggr ] z_Q(\la,m_Q^{\,\rm pole})\langle\qq\rangle_L\sim \frac{\mu^{\,\rm pole}_{gl}}{\la}\,\langle\qq\rangle_L\,,
\eeq
\bq
\nu_Q=\frac{3N_c-2N_F}{\nd}=\gamma_Q-1\,<\,-1\,,\quad \mu^{\,\rm pole}_{gl}\sim\frac{\langle\qq\rangle_L}{\la}\sim m_Q^{\,\rm pole}\sim\Biggl (\frac{\mph}{\la}\Biggr )^{\frac{\nd}{N_F-2N_c}} \gg\la\,. \label{(3.4)}
\eq
Because $N_F>3N_c$ and the global flavor symmetry is unbroken in these L-vacua, the rank restrction implies that quarks are not higgsed (as otherwise the flavor symmetry would be broken spontaneously), and so {\bf the overall phase is HQ} (heavy quarks).

The running mass of $\Phi$ looks as, see \eqref{(1.4)},
\bq
\mu_{\Phi}(\mu\gg\la)=\frac{\mph}{z_{\Phi}(\la,\mu)}\,,\quad z_{\Phi}(\la,\mu)=\Bigl (\frac{\mu}{\la}\Bigr )^{\gamma_{\Phi}}\gg 1\,,\quad 1\,<\gamma_{\Phi}=-2\gamma_Q\,<\,2\,. \label{(3.5)}
\eq
Therefore, the running mass $\mu_{\Phi}(\mu)$ {\it decreases} with increasing $\mu>\la$ and (if nothing prevents) $\mu_{\Phi}(\mu>\mos)<\mu$, so that the field $\Phi$ {\it becomes relevant} at scales $\mu>\mos$, see section 4 in \cite{ch4} and Appendix in \cite{ch6},
\bq
\la\ll\mu_{o}^{str}=\la\Bigl (\frac{\la}{\mph}\Bigr )^{\frac{1}{2\gamma_Q-1}}=\la\Bigl (\frac{\mph}{\la}\Bigr )^{\frac{\nd}{3N_F-5N_c}}\ll\mph,\quad \frac{\mu_{o}^{str}}{m_Q^{\,\rm pole}}\sim\Bigl (\frac{\la}{\mph}\Bigr )^{\frac{\nd(2N_F-3N_c)}{(N_F-2N_c)(3N_F-5N_c)}}\ll 1\,. \label{(3.6)}
\eq
Thus, when we are going down from high energy, because $\mu_{o}^{str}\ll\mph\ll m_Q^{\,\rm pole}$, at the scale $\mu=m_Q^{\,\rm pole}$ where all quarks decouple as heavy, all fions are already relevant, i.e. their running mass $\mu_{\Phi}(\mu=m_Q^{\,\rm pole})\ll\mph\ll m_Q^{\,\rm pole}$. Because the RG evolution of all fions becomes frozen at $\mu<m_Q^{\,\rm pole}$, this means that there is no pole in propagators of fions at $\mu\sim\mph$ (and so $\mu_{\Phi}=\mph(\mu=\la)$ is
a formal mass parameter), the pole mass of all $N_F^2$ fions will be $\mu^{\rm pole}(\Phi)\sim\mu_{\Phi}(\mu=m_Q^{\,\rm pole})\ll\mph$, see \eqref{(3.9)} below.

All quarks decouple as heavy at $\mu<m_Q^{\,\rm pole}$ and there remains at this point the $SU(N_c)$ SYM in the strong coupling regime. The scale factor $\lambda_{YM}$ of its gauge coupling is determined from the matching, see \eqref{(1.4)},\eqref{(1.5)},\eqref{(3.2)},
\bq
a_{(+)}(\mu=m_Q^{\,\rm pole})=\Bigl (\frac{\la}{m_Q^{\,\rm pole}}\Bigr )^{\nu_Q}=a_{YM}(\mu=m_Q^{\,\rm pole})=\Bigl (\frac{m_Q^{\,\rm pole}}{\lambda_{YM}}\Bigr )^3\quad\ra\quad\lambda_{YM}=\langle\lym^{(L)}\rangle\ll m_Q^{\,\rm pole}\,,
\eq
as it should be, see \eqref{(3.1)}. Decreasing scale further and integrating out gluons at $\mu<\langle\lym^{(L)}\rangle$ via the VY procedure \cite{VY}, the Lagrangian at $\mu=\langle\lym^{(L)}\rangle$ looks as, see \eqref{(1.1)} for ${\cal W}_{\Phi}$,
\bbq
K=z_{\Phi}(\la,m_Q^{\,\rm pole})\,{\rm Tr\,}(\Phi^\dagger\Phi)\,,\quad  z_{\Phi}(\la,m_Q^{\,\rm pole})=\Bigl (\frac{m_Q^{\,\rm pole}}{\la} \Bigr )^{\gamma_{\Phi}=-2\gamma_Q}\sim\frac{\mph^2}{\la^2}\gg 1\,,
\eeq
\bq
\w={\cal W}_{\Phi}+N_c\Bigl (\la^{3N_c-N_F}\det  m_Q^{\rm tot} \Bigr )^{1/N_c}\,,\quad  m_Q^{\rm tot}=m_Q-\Phi\,. \label{(3.8)}
\eq
From \eqref{(3.8)}, the masses of all $N_F^2$ fions $\Phi^j_i$ are
\bq
\mu^{\,\rm pole}(\Phi)\sim\frac{\mph}{z_{\Phi}(\la,m_Q^{\,\rm pole})}\sim\frac{\la^2}{\mph}\ll\la\ll\langle\lym^{(L)}\rangle\, \label{(3.9)}
\eq
(both terms in the superpotential \eqref{(3.8)} give parametrically the same contributions).

Recall that there is no pole in the $\Phi$-propagator at $\mu>\la$. There is only one generation of all $N_F^2$ fions $\Phi^j_i$ in this case with small masses \eqref{(3.9)}.

\subsection{Dual Theory}

The dual $SU(\nd)$ theory with $N_F>3N_c$ quark flavors is UV free and logarithmically weakly coupled at scales $\mu>\la$. The potentially important masses look in these L-vacua as follows (ignoring below all logarithmic factors for simplicity). The masses of dual quarks, see \eqref{(3.1)},
\bq
\mu^{\,\rm pole}_q\sim\frac{\langle M\rangle_L}{\la}\sim\la\Bigl (\frac{\mph}{\la}\Bigr )^{\frac{\nd}{N_F-2N_c}}\sim m_Q^{\,\rm pole}\gg\mph\gg\langle\lym^L\rangle\gg\la\,. \label{(3.10)}
\eq
The masses of dual gluons due to possible higgsing of quarks
\bq
{\ov\mu}^{\,2}_{gl}\sim\langle{\ov q} q\rangle\sim\la\Bigl (\frac{\mph}{\la}\Bigr )^{\frac{N_c}{N_F-2N_c}}\gg\la\,,\quad \frac{{\ov\mu}_{gl}}{\mu^{\,\rm pole}_q}\sim
\Bigl (\frac{\la}{\mph}\Bigr )^{\frac{2N_F-3N_c}{2(N_F-2N_c)}}\ll 1\,. \label{(3.11)}
\eq
Therefore, the dual quarks are not higgsed and {\bf the overall phase is Hq} (heavy quarks). All dual quarks decouple as heavy at $\mu<\mu^{\,\rm pole}_q$ in the weak coupling regime and the weak logarithmic RG evolution of all $N_F^2$ mions $M^i_j$ becomes frozen. All $N_F^2$ fions $\Phi^j_i$ have large masses $\mu^{\,\rm pole}(\Phi)\sim\mph\gg\la$ and are irrelevant at $\mu\ll\mph$.

Integrating out all dual quarks as heavy at $\mu<\mu^{\,\rm pole}_q$, then all fions $\Phi^j_i$ at $\mu<\mph$, and finally all  gluons at $\mu<\langle\lym^{(L)}\rangle$ via the VY procedure \cite{VY}, the dual Lagrangian at $\mu=\langle\lym^L\rangle\gg\la$ looks as, see \eqref{(1.3)} for ${\cal W}_{M}$,
\bq
K={\rm Tr\,}\frac{M^\dagger M}{\la^2}\,,\quad \w={\cal W}_{M} +{\cal W}_{\rm non-pert}\,,\quad {\cal W}_{\rm non-pert}=-\nd \Bigl (\la^{3\nd-N_F}\det\frac{ M}{\la}\Bigr )^{1/\nd}\,. \label{(3.12)}
\eq
From \eqref{(3.12)}, the masses of all all $N_F^2$ mions $M_j^i$ are, see \eqref{(3.1)},
\bq
\mu^{\,\rm pole}(M)\sim \frac{\la^2\langle S\rangle_L}{\langle M\rangle^2_L}\sim\frac{\la^2}{\mph}\ll\la \label{(3.13)}
\eq
(both terms of $\w$ in \eqref{(3.12)} give parametrically the same contributions).

The overall mass hierarchies look here as follows
\bq
\mu^{\,\rm pole}(M)\ll\la\ll\langle\lym^{(L)}\rangle\ll\mu^{\,\rm pole}(\Phi)\ll\mu^{\,\rm pole}_q\sim m^{\,\rm pole}_Q\,. \label{(3.14)}
\eq

At scales $\la^2/\mph\ll\mu\ll\langle\lym^{(L)}\rangle$, there are only effectively massless $N_F^2$ fions $\Phi^j_i$ in the direct theory, see \eqref{(3.9)}, and $N_F^2$ mions $M^i_j$ in the dual one, see \eqref{(3.13)}. Therefore, the t' Hooft triangles $SU^{3}(N_F)_L$ have opposite signs in the direct and dual theories.\\

\vspace*{1mm}

\addcontentsline{toc}{section}
 {\hspace*{3cm} Broken flavor symmetry $U(N_F)\ra U(\no)\times U(\nt)$}
{\bf \Large\hspace*{2cm} Broken flavor symmetry $U(N_F)\ra U(\no)\times U(\nt)$}
\section{Lt vacua}

\subsection{Direct and dual theories}

The quark condensates are parametrically the same as in L-vacua, see Appendix, the difference is that $\langle\Qo\rangle_{\rm Lt}\neq\langle\Qt\rangle_{\rm Lt}$,
\bq
\frac{N_c-\no}{N_c}\langle\Qo\rangle_{\rm Lt}\simeq -\frac{N_c-\nt}{N_c}\langle\Qt\rangle_{\rm Lt}\sim\la^2\Biggl (\frac{\mph}{\la}\Biggr )^{\frac{\nd}{N_F-2N_c}}\gg\la^2\,. \label{(4.1)}
\eq
Therefore, all nonzero masses are also parametrically the same, in particular all $\no^2$ fions $\Phi^1_1$, $\no^2$ mions $M^1_1$,  $\nt^2$ fions $\Phi^2_2$, and  $\nt^2$ mions $M^2_2$ have masses $\sim\la^2/\mph\ll\la\ll\lym^{(L)}$, see \eqref{(3.9)},\eqref{(3.13)}. The qualitative difference with the L-vacua is that the flavor symmetry is broken spontaneously and here appear massless Nambu-Goldstone particles. These are the hybrid fions $\Phi^1_2,\,\Phi^2_1$ in the direct theory, and hybrid mions $M^1_2,\,M^2_1$ in the dual one. Therefore, as in L-vacua, the t' Hooft triangles $SU^{3}(N_F)_L$ at scales $\mu<\langle\lym^{(L)}\rangle$ also have opposite signs in the direct and dual theories in these Lt-vacua.

\section{Special vacua}

\subsection{Direct theory}

The condensates look in these vacua with $\no=N_c,\, \nt=\nd$ as, see Appendix,
\bq
\langle\Qt\rangle=\frac{N_c}{2N_c-N_F} m_Q\mph\,,\quad \langle\Qo\rangle=\la^2\Bigl (\frac{\mph}{\la} \Bigr )^{\frac{\nd}{N_F-2N_c}}
\gg\mph\gg\la^2\,, \label{(5.1)}
\eq
\bbq
\langle S\rangle=\frac{\langle\Qo\rangle\langle\Qt\rangle}{\mph}\sim m_Q\la^2\Bigl (\frac{\mph}{\la} \Bigr )^{\frac{\nd}{N_F-2N_c}}\,,\quad \frac{\langle\Qt\rangle}{\langle\Qo\rangle}
\sim\frac{m_Q}{\la}\Bigl (\frac{\la}{\mph}\Bigr )^{\frac{N_c}{N_F-2N_c}}\ll 1\,,
\eeq
\bbq
\langle m_{Q,2}^{\,\rm tot}\rangle=\frac{\langle\Qo\rangle}{\mph}=\la\Bigl (\frac{\mph}{\la}\Bigr )^{\frac{N_c}{N_F-2N_c}}\gg\la\,,\quad
\langle m_{Q,1}^{\,\rm tot}\rangle=\frac{\langle\Qt\rangle}{\mph}\sim m_Q\ll\la\,.
\eeq

Recall that this direct theory is logarithmically weakly coupled at scales $\mu\ll\la$ only, while it is strongly coupled at $\mu\gg\la,\, a(\mu\gg\la)\gg 1$.
The potentially most important masses look in these vacua as follows. The quark mass looks as, see \eqref{(1.4)},
\bq
m_{Q,2}^{\,\rm pole}=\frac{\langle m_{Q,2}^{\,\rm tot}\rangle}{z_Q(\la,m_{Q,2}^{\,\rm pole})}=\la\Bigl (\frac{\langle m_{Q,2}^{\,\rm tot}\rangle}{\la}\Bigr)^{{\frac{\nd}{N_c}}}\sim\frac{\langle\Qo\rangle}
{\la},\, z_Q(\la,\mu)=\Bigl (\frac{\mu}{\la} \Bigr )^{\gamma_Q},\, \gamma_Q=\frac{2N_c-N_F}{N_F-N_c}<0.\,\, \label{(5.2)}
\eq
The gluon mass due to possible higgsing of quarks looks as, see \eqref{(1.4)},\eqref{(5.1)},\eqref{(5.2)},
\bq
\Bigl (\mu^{\,\rm pole}_{gl,1}\Bigr )^2\sim a(\mu=\mu^{\,\rm pole}_{gl,1})\,z_Q(\la,\mu^{\,\rm pole}_{gl,1})
\langle\Qo\rangle\sim\frac{\mu^{\,\rm pole}_{gl,1}}{\la}\,\langle\Qo\rangle\,,\label{(5.3)}
\eq
\bbq
a(\mu=\mu^{\,\rm pole}_{gl,1})\sim \Bigl (\frac{\la}{\mu^{\,\rm pole}_{gl,1}}\Bigr )^{\nu_Q}\,,\quad
a(\mu^{\,\rm pole}_{gl,1}) z_Q(\la,\mu^{\,\rm pole}_{gl,1})\sim\frac{\mu^{\,\rm pole}_{gl,1}}{\la}\,,\quad \nu_Q=\gamma_Q-1=\frac{3N_c-2N_F}{N_F-N_c}<0,
\eeq
\bbq
\mu^{\,\rm pole}_{gl,1}\sim\frac{\langle\Qo\rangle}{\la}\equiv\frac{\langle\Pi_1\rangle}{\la}\sim\la\Bigl (\frac{\mph}{\la} \Bigr )^{\frac{\nd}{N_F-2N_c}}\sim m_{Q,2}^{\,\rm pole}\gg\mph\gg\la\,,\quad\mu^{\,\rm pole}_{gl,1}\gg \mu^{\,\rm pole}_{gl,2}\,,\quad m_{Q,2}^{\,\rm pole}\gg m_{Q,1}^{\,\rm pole}\,.
\eeq

Therefore, { \bf the overall phase is} $\mathbf{HQ_2-Higgs_1}$. The whole RG evolution becomes frozen at $\mu<m_{Q,2}^{\,\rm pole}\sim\mu^{\,\rm pole}_{gl,1}$.
Similarly to L-vacua in section 3.1, see \eqref{(3.2)},\eqref{(3.5)},\eqref{(3.6)}, the running mass of all $N_F^2$ fions $\Phi^j_i$ is $\mu_{\Phi}(\mu=m_{Q,2}^
{\,\rm pole})\ll m_{Q,2}^{\,\rm pole}$, i.e. they are all relevant at scales $\mu<m_{Q,2}^{\,\rm pole}$ and there is no pole in the fion propagators at $\mu\sim\mph$, all masses of fions are much smaller than $\mph$, see below.

After integrating $Q^2,{\ov Q}_2$ quarks as heavy at $\mu\sim m_{Q,2}^{\,\rm pole}$, there remain $SU(N_c)$ SYM and $N^\prime_F=N_c$ quarks $Q^1,{\ov Q}_1$ which are higgsed at the same scale: $\mu^{\,\rm pole}_{gl,1}\sim\mu_{Q,2}^{\,\rm pole}\sim\langle\Qo\rangle/\la$, see \eqref{(5.1)}. We use for this case with $N^\prime_F=N_c$ the form of superpotential proposed by Seiberg in ~\cite{S1}.

As for the Kahler term, we write it as
\bq
K={\rm Tr\,}\Biggl [z_{\Phi}(\la,m_{Q,2}^{\,\rm pole})\,\Phi^\dagger\Phi+\frac{(\Pi^1_1)^
{\dagger}\Pi^1_1}{\la^2}\,\Biggr ] +(B^\dagger_1 B_1+{\ov B}^{\,\dagger}_1\, {\ov B}_1)\,,\,\, \Pi^i_j=({\ov Q}_j Q^i)\,,\,\, i,j=1...N_c\,. \label{(5.4)}
\eq

The kinetic term $K_{\Pi}$ pions $\Pi_1^1$ in \eqref{(5.4)} needs some explanations. There are two different contributions to $K_{\Pi}$. The first one originates directly from the kinetic term of higgsed quarks, see \eqref{(5.2)},\eqref{(5.3)},
\bq
K_Q= z_Q^{+}(\la,\mgo){\,\rm Tr}\,\Biggl [\,(Q^{\,1}_1)^\dagger Q^1_1+ (Q^1_1\ra {\ov Q}^{\,1}_1)\,\Biggr ]\ra K^{\rm (Born)}_{\Pi}\sim z_Q^{+}(\la,\mgo)\,{\rm Tr}\,\sqrt{(\Pi^1_1)^\dagger\Pi^1_1}\,\,.\label{(5.5)}
\eq
The second one originates from the loop of either massive gluons or massive higgsed quarks (superpartners of massive gluons) integrated over the non-parametric interval of momenta $p_E\sim \mgt$. It looks parametrically as ( $\mgo(\Pi^1_1)^2\sim {(\Pi^1_1)^\dagger\Pi^1_1}/\la^2\,,$ see the last line in \eqref{(5.3)}, $\,{\textsf Q}^1_1,\,{\ov {\textsf Q}}^{\,1}_1$ are canonically normalized quark fields ):
\bbq
K^{\rm (loop)}_{\Pi}\sim z_Q^{+}(\la,\mgo)\,{\rm Tr}\,\,\langle\langle\, (Q^{\,1}_1)^\dagger Q^1_1\,\rangle\rangle={\rm Tr}\,\,\langle\langle\, ({\textsf Q}^{\,1}_1)^\dagger {\textsf Q}^1_1\,\rangle\rangle\sim
\eeq
\bq
\sim{\rm Tr}\,\int \frac{d^{\,4} p_E}{[\, p_{E}^2+\mgo(\Pi^1_1)^2\,]}\sim
 {\rm Tr}\, \mgo(\Pi^1_1)^2\,\,,\quad r=\frac{K^{\rm(loop)}_{\Pi}}{K^{\rm (Born)}_{\Pi}}\sim \,\, a_{+}(\mu=\mgo)\,.\label{(5.6)}
\eq
So, $r\ll 1$ if quarks were higgsed in the weak coupling regime $a_{+}(\mu=\mgo)\ll 1$, but $r\gg 1$ in our
case of the strong coupling regime $a_{+}(\mu=\mgo)\gg 1$. Therefore, $K_{\Pi}\simeq K^{\rm (loop)}_{\Pi}$ is as in \eqref{(5.4)}.

As for the superpotential, we write it as
\bbq
\w={\cal W}_{\Phi}+{\rm Tr\,} (m_{Q,1}^{\,\rm tot}\Pi^1_1)- {\rm Tr\,}\Bigl (\Phi^2_1\frac{\Pi^1_1}{m_{Q,2}^{\,\rm tot}}\Phi^1_2\Bigr )+{\cal W}_{\rm non-pert}\,,
\eeq
\bq
{\cal W}_{\rm non-pert}=A\Biggl (1-\frac{\langle m_{Q,2}^{\,\rm tot}\rangle}{m_{Q,2}^{\,\rm tot}}\det\Bigl (\frac{\Pi^1_1}{\lambda^2}\Bigr )+\frac{{\ov B}_1 B_1}{\lambda^2}\Biggr )\,,\quad \lambda^2=\langle\Qo\rangle=\langle\Pi_1\rangle\,,\quad \langle A\rangle=\langle S\rangle \,. \label{(5.7)}
\eq
In \eqref{(5.7)}: $A$ is the Lagrange multiplier field, the additional factor $\langle m_{Q,2}^{\,\rm tot}\rangle/m_{Q,2}^{\,\rm tot}$ in ${\cal W}_{\rm non-pert}$ is needed to obtain from \eqref{(5.7)} the right value of $\langle\Phi_2\rangle$, the third term in $\w$ originated from integrating out $\nd^{\,2}$ heavy quarks $Q^2,{\ov Q}_2$.

We obtain from \eqref{(5.4)},\eqref{(5.7)}:\\
a) the masses of $\nd^{\,2}$ fions $\Phi^2_2$ are, see \eqref{(5.1)},\eqref{(5.2)},
\bq
\mu^{\rm pole}(\Phi^2_2)\sim\frac{\mph}{z_{\Phi}(\la,m_{Q,2}^{\,\rm pole})}\sim\frac{\la^2}{\mph}\ll\la,\,\,\, z_{\Phi}(\la,m_{Q,2}^{\,\rm pole})=\Bigl (\frac{m_{Q,2}^{\,\rm pole}}{\la}\Bigr )^{\gamma_{\Phi}}\sim\frac{\mph^2}{\la^2}\,,\,\, \gamma_{\Phi}=-2\gamma_Q> 0\,, \label{(5.8)}
\eq
(the main contribution to $\mu^{\rm pole}(\Phi^2_2)$ originates from the term ${\cal W}_{\Phi}$ in \eqref{(5.7)}, see \eqref{(1.1)}\,;\\

b) $N_c^2$ fields $\Pi^1_1$ and $N_c^2$ fields $\Phi_1^1$ are mixed significantly and physical fields ${\wh\Pi}^{1}_1,\, {\wh\Phi}^{1}_1$ have masses
\bq
\mu^{\rm pole}({\wh\Pi}^{1}_1)\sim \mu^{\rm pole}({\wh\Phi}^{1}_1)\sim\frac{\la^2}{\mph}\ll\la\,; \label{(5.9)}
\eq
c) the baryons $B_1,\,{\ov B}_1$ have masses
\bq
\mu^{\rm pole}(B_1)=\mu^{\rm pole}({\ov B}_1)\sim\frac{\langle S\rangle}{\langle\Qo\rangle}=\langle m_{Q,1}^{\,\rm tot}\rangle\sim m_Q\ll\la\,; \label{(5.10)}
\eq
d) $2\no\nt$ hybrid fions $\Phi^1_2,\,\Phi^2_1$ are the Nambu-Goldstone particles and are massless.

\subsection{Dual Theory}

This $SU(\nd)$ theory is UV free at $N_F>3N_c$ and weakly coupled at $\mu>\la$, but it is strongly coupled at $\mu<\la$. The potentially most important masses look here as (ignoring all logarithmic factors in the weak coupling region for simplicity), see \eqref{(5.1)},\eqref{(1.7)},
\bq
\mu^{\rm pole}_{q,1}\sim\mu_{q,1}\equiv\mu_{q,1}(\mu=\la)\sim\frac{\langle M_1\rangle=\langle\Qo\rangle}{\la}\sim\la\Bigl (\frac{\mph}{\la}\Bigr )^{\frac{\nd}{N_F-2N_c}}\gg\mph\gg\la\,,\label{(5.11)}
\eq
\bbq
{\ov \mu}^{\,\rm pole}_{gl,2}\sim\langle ({\ov q} q)_2\rangle^{1/2}=\Bigl (\la\langle m_{Q,2}^{\,\rm tot}\rangle\Bigr )^{1/2}\sim\la\Bigl (\frac{\mph}{\la}\Bigr )^{\frac{N_c}{2(N_F-2N_c)}}\,,\quad \la\ll{\ov \mu}^{\,\rm pole}_{gl,2}\ll\mph\ll \mu^{\rm pole}_{q,1}\,,
\eeq
\bbq
\mu_{q,2}\sim\frac{m_Q\mph}{\la}\,,\quad \mu^{\rm pole}_{q,2}\sim\mu_{q,2}\gg\la\quad{\rm at}\,\,\,m_Q\mph>\la^2,
\quad \mu^{\rm pole}_{q,2}\sim\la\Bigl (\frac{m_Q\mph}{\la^2}\Bigr )^{\frac{N_c}{\nd}}\ll\la\quad{\rm at}\,\,\,{m_Q\mph<\la^2}.
\eeq
\bq
\frac{\mu^{\rm pole}_{q,2}}{{\ov \mu}^{\,\rm pole}_{gl,2}}\ll 1\quad{\rm at}\quad\frac{m_Q\mph}
{\la^2}<1,\quad\quad\frac{\mu^{\rm pole}_{q,2}}{{\ov \mu}^{\,\rm pole}_{gl,2}}\sim\Bigl (\frac{\mph}
{{\wh\mu}_{\Phi}}\Bigr)^{{\frac{2N_F-5N_c}{2(N_F-2N_c)}}}\quad{\rm at}\quad\frac{m_Q\mph}{\la^2}>1,\,\,
\label{(5.12)}
\eq
\bbq
{\wh\mu}_{\Phi}=\la\Bigl (\frac{\la}{m_Q}\Bigr )^{\frac{2(N_F-2N_c)}{2N_F-5N_c}}\gg\frac{\la^2}{m_Q}.
\eeq

As it is seen from \eqref{(5.12)}, the hierarchy $\mu^{\rm pole}_{q,2} \lessgtr {\ov\mu}^{\,\rm pole}_{gl,2}$ changes at $\mph \lessgtr {\wh\mu}_{\Phi}$. As a result, {\bf the overall phase changes also}: it is $\mathbf{Hq_1-Higgs_2}$ at  $\mph < {\wh\mu}_{\Phi}$  and $\mathbf{Hq_1-Hq_2}$ at  $\mph > {\wh\mu}_{\Phi}$.\\

{\bf A)}.\,\,  Consider first the region $\la\ll\mph\ll {\wh\mu}_{\Phi}$. The dual quarks $q_1, {\ov q}^1$ decouple as heavy in the weak coupling regime at $\mu=\mu^{\rm pole}_{q,1}\gg\mph\gg\la$ and there remains $SU(\nd)$ with $\nt=\nd$ flavors of quarks $q_2, {\ov q}^{\,2}$. The new scale factor of the gauge coupling is
\bq
\Bigl (\wh\Lambda\Bigr )^{3\nd-\nt=2\nd}=\la^{3\nd-N_F}\mu^{N_c}_{q,1}\,,\quad \wh\Lambda\sim \la\Bigl (\frac{\mph}{\la}\Bigr )^{\frac{N_c}{2(N_F-2N_c)}}\sim {\ov \mu}^{\,\rm pole}_{gl,2}\gg\la. \label{(5.13)}
\eq

After integrating out $q_1, {\ov q}^{\,1}$ quarks as heavy at $\mu<\mu^{\rm pole}_{q,1}$ and then all $N_F^2$ fions $\Phi$ at $\mu<\mu^{\,\rm pole}(\Phi)\sim\mph\gg\wh\Lambda$, the dual Lagrangian looks as (all logarithmic factors in the Kahler term are ignored for simplicity), see \eqref{(1.3)} for ${\cal W}_M$,
\bbq
K={\rm Tr\,} \Bigl [\frac{M^{\dagger} M}{\la^2}+(q_2)^{\dagger} q_2 +(q_2\ra {\ov q}^{\,2}) \Bigr ]\,,
\eeq
\bq
\w={\cal W}_M + {\rm Tr}\,\Biggl ({\ov q}^{\,2} M^1_2\frac{1}{\la M^1_1} M^2_1 q_2\Biggr ) -{\rm Tr}\,\Bigl ({\ov q}^{\,2}\frac{M^2_2}{\la} q_2\Bigr )\,, \label{(5.14)}
\eq
\bbq
{\cal W}_M=m_Q {\rm Tr}\,(M)-\frac{1}{2\mph}\Biggl ({\rm Tr}(M^2)-\frac{1}{N_c}({\rm Tr} M)^2 \Biggr ).
\eeq

All $\nd$ flavors of quarks $q_2, {\ov q}^{\,2}$ are higgsed at $\mu\sim {\ov \mu}^{\,\rm pole}_{gl,2}\sim{\wh\Lambda}\ll\mph\ll\mu^{\rm pole}_{q,1}$ (already in the strong coupling region ${\ov a}(\mu\sim{\wh\Lambda})\sim 1$), and we use for this case with $N_F^\prime=\nt=\nd$ the form  of ${\cal W}_{\rm non-pert}$ proposed in \cite{S1},
\bbq
K={\rm Tr\,} \Bigl [\frac{M^{\dagger} M}{\la^2}+ 2\sqrt{(N^2_2)^{\dagger} N^2_2}\,\Biggr ] +\Bigl (B^{\dagger}_2 B_2+ (B_2\ra {\ov B}_2)\,\Bigr )\,,\quad N^j_i=({\ov q}^j q_i)\,,\,\, i,j=N_c+1...N_F\,,
\eeq
\bq
\w={\cal W}_M - {\rm Tr\,}\,\frac{M^2_2 N^2_2}{\la} +{\rm Tr\,}\,\Biggl ( M^1_2\frac{N^2_2}{\la M^1_1} M^2_1\Biggr )+{\cal W}_{\rm non-pert}\,, \label{(5.15)}
\eq
\bq
{\cal W}_{\rm non-pert}={\ov A}\Biggl (1-\det\Bigl (\frac{N^2_2}{\langle N_2\rangle}\Bigr )+\frac{{\ov B}_2 B_2}{\langle N_2\rangle} \Biggr ),\,\, \langle{\ov A}\rangle=\langle {\ov S}\rangle=\langle -S\rangle,\,\,\langle N_2\rangle\sim (\wh\Lambda)^2\sim \la^2\Bigl (\frac{\mph}{\la}\Bigr )_{,}^{\frac{N_c}{N_F-2N_c}}\,\,\, \label{(5.16)}
\eq
where ${\ov A}$ is the Lagrange multiplier field and $N^2_2$ are $\nd^{\,2}$ dual pions (nions).

We obtain from \eqref{(5.15)},\eqref{(5.16)} for the particle masses\,:\\

a)  $\nd^{\,2}$ fields $M^2_2$ and $\nd^{\,2}$ fields $N^2_2$  have masses
\bq
\la\ll\mu^{\,\rm pole}( M^{\,2}_2)\sim\mu^{\,\rm pole}( N^{\,2}_2)\sim{\ov \mu}^{\,\rm pole}_{gl,2}\sim\la\Bigl (\frac{\mph}{\la}\Bigr )^{\frac{N_c}{2(N_F-2N_c)}}\ll\mph\,, \label{(5.17)}
\eq
(the main contribution originates from the second term in the superpotential \eqref{(5.15)}\,),\\

b) the masses of baryons are
\bq
\mu^{\,\rm pole}(B_2)=\mu^{\,\rm pole}({\ov B}_2)\sim\frac{\langle M_2\rangle}{\la}\sim\frac{m_Q\mph}{\la}\ll\mu^{\,\rm pole}( M^{\,2}_2)\,, \label{(5.18)}
\eq

c) $N_c^2$ mions $M^1_1$ have masses
\bq
\mu^{\,\rm pole}(M^1_1)\sim \frac{\la^2}{\mph}\ll\la\,, \label{(5.19)}
\eq
d) $2\no\nt=2\nd N_c$ hybrid mions $M^1_2,\, M^2_1$ are the Nambu-Goldstone particles and are massless.

The hierarchies of nonzero masses look in this region $\la\ll\mph\ll{\wh\mu}_{\Phi}$ as
\bq
\mu^{\,\rm pole}(M^1_1)\ll\la\ll\mu^{\,\rm pole}( M^{\,2}_2)\sim\mu^{\,\rm pole}( N^{\,2}_2)\sim{\ov \mu}^{\,\rm pole}_{gl,2}\ll\mph\ll\mu^{\rm pole}_{q,1}\,,\quad \mu^{\,\rm pole}(M^1_1)\lessgtr\mu^{\,\rm pole}(B_2)\,. \label{(5.20)}
\eq
\vspace{1mm}

{\bf B)}\,\,  The region $\mph\gg {\wh\mu}_{\Phi}$. The Lagrangian \eqref{(5.14)} remains the same, the difference at lower energies is that quarks ${\ov q}^{\,2}, q_2$ are not higgsed now but decouple as heavy at $\mu<\mu^{\rm pole}_{q,2}\sim m_Q\mph/\la,\,\mu^{\rm pole}_{q,2}\gg{\wh\Lambda}$, still in the weak coupling regime, see \eqref{(5.12)},\eqref{(5.13)}. The scale factor of the gauge coupling of remained unbroken $SU(\nd)$ SYM at $\mu<\mu^{\rm pole}_{q,2}$ looks as, see \eqref{(5.1)},
\bq
\lambda_{YM}^3=\la^{3\nd-N_F}\mu_{q,1}^{N_c}\,\mu_{q,2}^{\nd}\,,\quad \langle\lambda_{YM}^3\rangle=m_Q\la^2\Bigl (\frac{\mph}{\la}\Bigr )^{\frac{\nd}{N_F-2N_c}}=\langle S\rangle\,, \label{(5.21)}
\eq
as it should be. The hierarchies of scales look as: $\mu^{\,\rm pole}_{q,1}\gg\mph\gg\mu^{\,\rm pole}_{q,2}
\gg\langle\lym^{(\rm spec)}\rangle\gg\la$. All $N_F^2$ fion fields $\Phi$ with masses $\mu^{\,\rm pole}(\Phi)\sim\mph$ can also be integrated out at scales $\mu<\mph$. Therefore, after integrating out all quarks and all $\Phi$ as heavy and, finally, $SU(\nd)$ gluons at $\mu=\langle\lym^{(\rm spec)}\rangle$ via the VY procedure \cite{VY}, the Lagrangian looks as, see \eqref{(5.14)} for ${\cal W}_M$,
\bq
K={\rm Tr\,}\frac{M^{\dagger} M}{\la^2}\,,\quad \w={\cal W}_M +{\cal W}_{\rm non-pert}\,,\quad {\cal W}_{\rm non-pert}=-\nd\Bigl (\la^{3\nd-N_F}\det\frac{M}{\la}\Bigr )^{1/\nd}\,. \label{(5.22)}
\eq
From \eqref{(5.22)}, the masses of $N^2_c$ mions $M^1_1$ and $\nd^{\,2}$ mions $M^2_2$ are (up to logarithmic factors), see \eqref{(5.1)},
\bq
\mu^{\,\rm pole}(M^1_1)\sim\frac{\la^2}{\mph}\ll\mu^{\,\rm pole}(M^2_2)\sim\frac{\langle\Qo\rangle}{\langle\Qt\rangle}
\frac{\la^2}{\mph}\sim\frac{\la^2}{m_Q}\Bigl (\frac{\la}{\mph}\Bigr )^{\frac{N_F-3N_c}{N_F-2N_c}}\ll\langle\lym^{(\rm spec)}\rangle\,,\quad\mph\gg {\wh\mu}_{\Phi}\,, \label{(5.23)}
\eq
(the main contribution to $\mu^{\,\rm pole}(M^1_1)$ in \eqref{(5.23)} originates from ${\cal W}_M$, while $\mu^{\,\rm pole}(M^2_2)$ is dominated by the contribution from ${\cal W}_{\rm non-pert}$ ).

$2\nd N_c$ hybrid mions $M^1_2, M^2_1$ are the Numbu-Goldstone particles and are massless.

The hierarchies of nonzero masses look in this region $\mph\gg{\wh\mu}_{\Phi}$ as
\bq
\mu^{\,\rm pole}(M^1_1)\ll\mu^{\,\rm pole}(M^2_2)\ll\langle\lym^{(\rm spec)}\rangle\ll\mph\ll\mu^{\rm pole}_{q,1}\,.
\eq

\section{br1 vacua, $\,\la\ll\mph\ll\la^2/m_Q$}

\subsection{Direct theory}

The condensates look in these vacua with $\no<N_c, \nt>\nd$ as, see Appendix,
\bq
\langle\Qo\rangle_{\rm br1}\simeq \frac{N_c}{N_c-\no} m_Q\mph,\quad \langle\Qt\rangle_{\rm br1}\sim \la^2\Bigl (\frac{\la}{\mph}\Bigr )^{\frac{\no}{N_c-\no}}
\Bigl (\frac{m_Q}{\la}\Bigr )^{\frac{\nt-N_c}{N_c-\no}}\,, \label{(6.1)}
\eq
\bbq
\langle\lym^{(\rm br1)}\rangle^3\equiv\langle S\rangle_{\rm br1}\sim\la^3\Bigl (\frac{\la}{\mph}\Bigr )^{\frac{\no}{N_c-\no}}\Bigl (\frac{m_Q}{\la}\Bigr )^{\frac{\nt-\no}{N_c-\no}}\,,\quad \frac{\lym^{(\rm br1)}}{m_Q}\ll 1\,,\quad \frac{\langle\Qt\rangle_{\rm br1}}{\langle\Qo\rangle_{\rm br1}}\ll 1\,.
\eeq

This direct $SU(N_c)$ theory is in the weak coupling IR free logarithmic regime at scales $\mu<\la$. We consider in this section 6 only the region $m_Q\mph\ll\la^2$.

All $N_F^2$ fions $\Phi^j_i$ have masses $\la\ll\mu^{\rm pole}(\Phi)\sim\mos\ll\mph$, see section 2.1 and \eqref{(2.4)}, and become too heavy and irrelevant at $\mu<\mos$. Therefore, they all can be integrated out as heavy at $\mu<\la$, but it is technically more convenient to retain them all as irrelevant auxiliary fields.

The largest masses smaller than $\la$ have $\no(2N_c-\no)$ gluons due to higgsing of $Q^1, {\ov Q}_1$ quarks (all logarithmic factors are ignored for simplicity)
\bq
\mu^{\rm pole}_{gl,1}\sim \langle\Pi_1\rangle^{1/2}=\langle\Qo\rangle^{1/2}_{\rm br1}\sim (m_Q\mph)^{1/2}\ll\la\,. \label{(6.2)}
\eq
At scales $\mu<\mu^{\rm pole}_{gl,1}$ there remain $SU(N_c-\no)$ gauge group, $2\nt(N_c-\no)$ still active quarks ${\ov Q}^{\,\prime}_2,\, Q^{\,\prime,2}$  with unbroken colors, $2\no\nt$ hybrid pions $\Pi^1_2={\ov Q}_2\langle Q^1\rangle,\, \Pi^2_1=\langle {\ov Q}_1\rangle Q^2$ (in essence, these are quarks $Q^2, {\ov Q}_2$ with broken colors), and $\no^2$ pions $\Pi^1_1$.

The next mass scale $m^{\rm pole}_{Q,2}\ll\mu_{gl,1}$ is those of ${\ov Q}^{\,\prime}_2, Q^{\,\prime,2}$ quarks, see \eqref{(6.1)} and \eqref{(A.4)},
\bq
m^{\rm pole}_{Q,2}\sim \langle m^{\rm tot}_{Q,2}\rangle\sim m_Q\gg\mu^{\rm pole}_{gl,2}\sim\langle\Qt\rangle^{1/2}_{\rm br1}\,,\quad m^{\rm tot}_{Q,2}=m_Q-\Phi^2_2\,. \label{(6.3)}
\eq
Therefore, {\bf the overall phase is} $\mathbf{Higgs_1-HQ_2}$. After integrating out all higgsed quarks and gluons at $\mu<\mu^{\rm pole}_{gl,1}$ and then ${\ov Q}^{\,\prime}_2, Q^{\,\prime,2}$ quarks as heavy at $\mu<m^{\rm pole}_{Q,2}$, the scale factor $\lambda_{YM}$ of the $SU(N_c-\no)$ SYM is, see \eqref{(6.1)},
\bq
\lambda^{3(N_c-\no)}_{YM}=\frac{\la^{3N_c-N_F}\langle m^{\rm tot}_{Q,2}\rangle^{\nt}}
{\langle\det\Pi^1_1\rangle}\quad\ra\quad  \lambda_{YM}=\langle\lym^{(\rm br1)}\rangle\,, \label{(6.4)}
\eq
as it should be. Integrating then out $SU(N_c-\no)$ gluons via the VY procedure \cite{VY}, the Lagrangian at $\mu=\langle\lym^{\rm br1}\rangle$ looks as, see \eqref{(1.1)} for ${\cal W}_{\Phi}$, \eqref{(6.1)} and \eqref{(A.4)},
\bq
K={\rm Tr}\,(\Phi^\dagger\Phi)+  {\rm Tr}\,\Biggl [2\sqrt{(\Pi^1_1)^{\dagger}\Pi^1_1}+\Pi_2^1\frac{1}
{\sqrt{(\Pi^1_1)^{\dagger}\Pi^1_1}}\,(\Pi_2^1)^{\dagger}+(\Pi_1^2)^{\dagger}\frac{1}{\sqrt{
(\Pi^1_1)^{\dagger}\Pi^1_1}}\,\Pi_1^2 \Biggr ]\,, \label{(6.5)}
\eq
\bbq
{\cal W}={\cal W}_{\Phi}+{\rm Tr}\,\Bigl (m_{Q,1}^{\rm tot}\,\Pi^1_1\Bigr )+{\cal W}_{\rm hybr}+{\cal W}_{\rm non-pert}\,,\quad {\cal W}_{\rm hybr}={\rm Tr}\,\Bigl (\Pi_2^1\frac{m^{\rm tot}_{Q,2}\,}{\Pi^1_1}\Pi_1^2-\Phi_1^2\Pi_2^1-\Phi_2^1\Pi_1^2\Bigr )\,,
\eeq
\bbq
{\cal W}_{\rm non-pert}=(N_c-\no)\Bigl (\,\lym^{(\rm br1)}\,\Bigr )^3= (N_c-\no)\Biggl [\frac{\la^{3N_c-N_F}\,\det m^{\rm tot}_{Q,2}}{\det \Pi^1_1}\Biggr ]^{1/(N_c-\no)}\,.
\eeq
The masses of $\no^{\,2}$ pions $\Pi^1_1$ and $2\no\nt$ hybrids from \eqref{(6.5)} are
\bq
\mu^{\rm pole}(\Pi^1_1)\sim\frac{\langle\Pi^1_1\rangle=\langle\Qo\rangle_{\rm br1}}{\mph}\sim m_Q\sim m^{\rm pole}_{Q,2}\,,\quad
\langle\Pi^1_{1^\prime}\rangle=\langle({\ov Q}_{1^\prime} Q^1)\rangle_{\rm br1}= \delta^1_{1^\prime} \frac{N_c}{N_c-\no}\,m_Q\mph\,, \label{(6.6)}
\eq
\bbq
\mu^{\rm pole}(\Pi^1_2)=\mu^{\rm pole}(\Pi^2_1)=0\,,
\eeq
(the main contribution to $\mu^{\rm pole}(\Pi^1_1)$ gives the term $\sim (\Pi^1_1)^2/\mph$ originating from  ${\rm Tr}\, (m^{\rm tot}_{Q,1}\,\Pi^1_1 )$ in \eqref{(6.5)} after integrating out heavy $\Phi^1_1$ with masses $\gg\la\,$ ).\\

On the whole, the mass spectrum of this ${\cal N}=1\,\,SU(N_c)$ theory look in this case as follows.\\
1) There are $\no(2N_c-\no)\,\,{\cal N}=1$ multiplets of massive gluons, $\mu^{\rm pole}_{\rm gl,1}\sim (m_Q\mph) ^{1/2}\ll\la$.\\
2) A large number of hadrons made from nonrelativistic and weakly confined ${\ov Q}^{\,\prime}_2,\,Q^{\,2,\,\prime}$ quarks with $\nt>N_c$ flavors and $N_c-\no$ colors, their mass scale is $\mu_H\sim m_Q,\,\, \langle\lym^{(\rm br1)}\rangle\ll \mu_H\ll\mu_{\rm gl,1}$ (the tension of the confining string originating from $SU(N_c-\no)$ SYM is $\sigma^{1/2}\sim\langle\lym^{(\rm br1)}\rangle\ll m_Q$).\\
3) $\no^2$ pions $\Pi^1_1$ with masses $\sim m_Q$.\\
4) A large number of ${\cal N}=1\,\, SU(N_c-\no)$ SYM strongly coupled gluonia with the mass scale $\sim\langle\lym^{(\rm br1)}\rangle$, see \eqref{(6.1)}.\\
5) $2\no\nt$ massless Nambu-Goldstone ${\cal N}=1$ multiplets $\Pi^1_2,\,\Pi^2_1$.\\
6) All $N_F^2$ fields $\Phi$ have masses $\la\ll\mu^{\rm pole}(\Phi)\sim\mos\ll\mph$, see \eqref{(2.4)}, and are dynamically irrelevant at lower scales.

\subsection{Dual Theory}
The condensates in this dual theory \eqref{(1.2)} at the scale $\mu=\la$ look as, see \eqref{(6.1)},
\bq
\langle M_i\rangle=\langle ({\ov Q}Q)_i\rangle,\,\,i=1,2,\quad \langle ({\ov q}q)_2\rangle=\frac{\la\langle M_1\rangle}{\mph}\sim m_Q\la\,,\quad
\frac{\langle ({\ov q}q)_1\rangle}{\langle ({\ov q}q)_2\rangle}=\frac{\langle\Qt\rangle}{\langle\Qo\rangle}\ll 1\,. \label{(6.7)}
\eq

Recall that we consider only the region $m_Q\mph\ll\la^2$ and this dual theory is strongly coupled at scales $\mu<\la$. All masses of dual quarks and gluons are $\ll\la$ in this case, so that we have to deal with the strong coupling regime of this theory, ${\ov a}(\mu\ll\la)=(\la/\mu)^{\nu_q>\,0}\gg 1$. All $N_F^2$ fions $\Phi^j_i$ have masses $\mu^{\rm pole}(\Phi)\sim\mph\gg\la$, they are too heavy and dynamically irrelevant at $\mu<\mph$. Therefore, they all can be integrated out from the beginning, see \eqref{(1.3)}. The potentially important masses look as follows:\\
a) the masses of dual quarks
\bq
\mu_{q,1}\equiv\mu_{q,1}(\mu=\la)=\frac{\langle M_1\rangle}{\la}\sim \frac{m_Q\mph}{\la}\,,\,\, \mu^{\,\rm pole}_{q,1}=\frac{\mu_{q,1}}{z^{(+)}_q(\la,\mu^{\,\rm pole}_{q,1})}\sim\la\Bigl ( \frac{m_Q\mph}{\la^2}\Bigr )^{N_c/\nd}\ll\la\,, \label{(6.8)}
\eq
\bq
\mu_{q,2}=\frac{\langle M_2\rangle}{\la},\quad\mu^{\,\rm pole}_{q,2}=\frac{\mu_{q,2}}{z^{(+)}_{q}(\la, \mu^{\,\rm pole}_{q,1})z^{(-)}_{q}( \mu^{\,\rm pole}_{q,1},\, \mu^{\,\rm pole}_{q,2})}\sim m_Q\ll\mu^{\,\rm pole}_{q,1},\,\,\, z^{(\pm)}_{q}(\mu_1,\mu_2)=\Bigl (\frac{\mu_2}{\mu_1}\Bigr )_{,}^{\gamma_q^{(\pm)}}\,\, \label{(6.9)}
\eq
\bbq
\gamma_q^{(+)}=\frac{N_F-2N_c}{N_c}\,,\,\,\nu_q^{(+)}=\frac{N_F-3N_c}{N_c},\,\,\gamma_q^{(-)}=
\frac{N_F-2N_c+\no}{N_c-\no}\,,\,\,\nu_q^{(-)}=\frac{N_F-3N_c+2\no}{N_c-\no}\,,\quad \gamma_{\Phi}=0\,;
\eeq
b) the gluon masses due to possible higgsing of dual quarks, see \eqref{(6.7)}-\eqref{(6.9)},
\bq
\Bigl ({\ov\mu}^{\,\rm pole}_{gl,2}\Bigr )^2\sim\rho^{(+)}\rho^{(-)}\,\langle ({\ov q}q)_2\rangle\sim\frac{{\ov\mu}^{\,\rm pole}_{gl,2}}{\la}\,\langle ({\ov q}q)_2\rangle,\,\, {\ov\mu}^{\,\rm pole}_{gl,2}\sim m_Q\gg{\ov\mu}^{\,\rm pole}_{gl,1},\,\, {\ov a}(\mu=\mu^{\,\rm pole}_{q,1})\sim\Bigl (\frac{\la}{\mu^{\,\rm pole}_{q,1}}\Bigr )_{,}^{\nu_{q}^{(+)}} \label{(6.10)}
\eq
\bbq
\rho^{(+)}={\ov a}(\mu=\mu^{\,\rm pole}_{q,1})\,z^{(+)}_{q}(\la,\mu^{\,\rm pole}_{q,1})\sim\frac{\mu^{\,\rm pole}_{q,1}}{\la}\,,\quad \rho^{(-)}\sim\Bigl (\frac{\mu^{\,\rm pole}_{q,1}}{{\ov\mu}^{\,\rm pole}_{gl,2}}\Bigr )^{\nu_{q}^{(-)}} z_q^{(-)}(\mu^{\,\rm pole}_{q,1},{\ov\mu}^{\,\rm pole}_{gl,2})\sim\frac{{\ov\mu}^{\,\rm pole}_{gl,2}}{\mu^{\,\rm pole}_{q,1}}\,.
\eeq
Because $\no<N_c,\nt>\nd$ in these br1-vacua and the flavor symmetry is broken as $U(N_F)\ra U(\no)\times U(\nt)$, the rank condition implies that quarks $q_2,{\ov q}^{\,2}$ are not higgsed (as otherwise the flavor symmetry $U(\nt)$ will be {\it additionally} broken), i.e. $\mu^{\,\rm pole}_{q,2}=(\rm several)\,{\ov\mu}^{\,\rm pole}_{gl,2}$. Therefore, {\bf the overall phase is} $\mathbf{Hq_1-Hq_2}$ (heavy quarks).

Integrating then out $q_1, {\ov q}^1$ quarks as heavy at $\mu<\mu^{\,\rm pole}_{q,1}$, and then $q_2, {\ov q}^2$ at $\mu=\mu^{\,\rm pole}_{q,2}$, the scale factor of the remained $SU(\nd)$ SYM in the strong coupling regime is determined from the matching, see \eqref{(1.4)},\eqref{(1.5)},\eqref{(6.8)},\eqref{(6.9)},
\bq
{\ov a}(\mu=\mu^{\,\rm pole}_{q,2})=\Bigl (\frac{\la}{\mu^{\,\rm pole}_{q,1}}\Bigr )^{\nu^{(+)}_{q}}\Bigl (\frac{\mu^{\,\rm pole}_{q,1}}{\mu^{\,\rm pole}_{q,2}}\Bigr )^{\nu^{(-)}_{q}}={\ov a}_{YM}(\mu=\mu^{\,\rm pole}_{q,2})=\Bigl (\frac{\mu^{\,\rm pole}_{q,2}}{\lambda_{YM}} \Bigr )^3\quad\ra\quad
\lambda_{YM}=\langle\lym^{(\rm br1)}\rangle\,, \label{(6.11)}
\eq
as it should be, see \eqref{(6.1)}. Finally, integrating at $\mu<\langle\lym^{(\rm br1)}\rangle$ all $SU(\nd)$ gluons via the VY procedure \cite{VY}, the Lagrangian takes the form, see Appendix in \cite{ch6} and \eqref{(6.9)} for the anomalous dimensions,
\bq
K=\frac{z^{(+)}_M}{\la^2}\,{\rm Tr\,}\Bigl [\,(M^1_1)^{\dagger} M^1_1+(M^1_2)^{\dagger} M^1_2+(M^2_1)^{\dagger} M^2_1+z^{(-)}_M (M^2_2)^{\dagger} M^2_2 \Bigr ]\,, \label{(6.12)}
\eq
\bbq
z^{(+)}_M=z^{(+)}_M(\la,\mu^{\,\rm pole}_{q,1})=\Bigl (\frac{\la}{\mu^{\,\rm pole}_{q,1}}\Bigr )^{2\gamma^{(+)}_q}\,,\quad z^{(-)}_M=z^{(-)}_M(\mu^{\,\rm pole}_{q,1},\mu^{\,\rm pole}_{q,2})=\Bigl (\frac{\mu^{\,\rm pole}_{q,1}}{\mu^{\,\rm pole}_{q,2}}\Bigr )^{2\gamma^{(-)}_q}\,,
\eeq
\bq
\w={\cal W}_{M}+{\cal W}_{\rm non-pert}\,,\quad {\cal W}_{\rm non-pert}=-\nd\Bigl (\la^{3\nd-N_F}\det\frac{M}{\la} \Bigr )^{1/\nd}\,,\label{(6.13)}
\eq
\bbq
{\cal W}_{M}= m_Q{\rm Tr\,}(M) -\frac{1}{2\mph}\Biggl [{\rm Tr\,}(M^2)-\frac{1}{N_c}\Bigl ({\rm Tr}\,M \Bigr )^2 \Biggr ]\,.
\eeq

We obtain from \eqref{(6.12)},\eqref{(6.13)} for the masses of $N_F^2$ mions $M^i_j$\,:\\
a) masses of $\no^2$ mions $M^1_1$
\bbq
\mu^{\,\rm pole}(M^1_1)\simeq\frac{\la^2}{z^{(+)}_M(\la,\mu^{\,\rm pole}_{q,1})\mph}\sim\la\Bigl (\frac{m_Q}{\la}\Bigr )^{2(N_F-2N_c)/\nd}\Bigl (\frac{\mph}{\la}\Bigr )^{(N_F-3N_c)/\nd}\ll m_Q\,\,{\rm at}\,\, \mph\ll\frac{\la^2}{m_Q}\,,
\eeq
\bq
\frac{\mu^{\,\rm pole}(M^1_1)}{\mu^{\,\rm pole}_{q,1}}\sim\Biggl [\frac{m_Q}{\la}\Bigl (\frac{\mph}{\la}
\Bigr )^{\frac{N_F-4N_c}{2N_F-5N_c}}\Biggr ]^{(2N_F-5N_c)/\nd}\ll 1\quad {\rm at}\,\quad \mph\ll\frac{\la^2}{m_Q}\,,\quad \nd=N_F-N_c\,,\label{(6.15)}
\eq
the main contribution to $\mu^{\,\rm pole}(M^1_1)$ originates from the term $\sim (M^1_1)^2/\mph$ in \eqref{(6.13)}\,;\\
b)  masses of $\nt^2$ mions $M^2_2$
\bbq
\mu^{\,\rm pole}(M^2_2)\simeq\frac{\langle M_1\rangle}{\langle M_2\rangle}\frac{\la^2}{z^{(+)}_M(\la,\mu^{\,\rm pole}_{q,1})z^{(-)}_M(\mu^
{\,\rm pole}_{q,1},\mu^{\,\rm pole}_{q,2})\mph}\sim\la\Bigl (\frac{m_Q}{\la}\Bigr )^{\frac{N_F-2N_c}{N_c-\no}}\Bigl (\frac{\la}{\mph}\Bigr )^{\frac{\no}{N_c-\no}}\ll\mu^{\,\rm pole}(M^1_1)\,,
\eeq
\bq
\Bigl [\frac{\mu^{\,\rm pole}(M^2_2)}{\langle\lym^{(\rm br1)}\rangle}\Bigr ]^3\sim\Bigl (\frac{m_Q}{\la}\Bigr )^{\frac{2(N_F-3N_c+\no)}{N_c-\no}}\Bigl (\frac{\la}{\mph}\Bigr )^{\frac{2\no}{N_c-\no}} \ll 1\,, \label{(6.16)}
\eq
the main contribution to $\mu^{\,\rm pole}(M^2_2)$ originates from the term ${\cal W}_{\rm non-pert}$ in \eqref{(6.13)}\,; \\
c) the mions $M^1_2$ and $M^2_1$ are massless.\\

On the whole, the mass spectrum of this ${\cal N}=1\,\,SU(\nd)$ dual theory at $\mph\ll\la^2/m_Q$ look in this case as follows. All quarks are not higgsed but weakly confined.\\
1) Among the masses smaller than $\la$ the largest masses have $q_1, {\ov q}^1$ quarks,\\ $\mu^{\,\rm pole}_{q,1}\sim\la (m_Q\mph/\la^2 )^{N_c/\nd}\ll\la$, see \eqref{(6.8)}.\\
2) The next mass scale is those of $q_2, {\ov q}^2$ quarks, $\mu^{\,\rm pole}_{q,2}\sim m_Q\ll \mu^{\,\rm pole}_{q,1}$, see \eqref{(6.9)}.\\
3) There is a large number of strongly coupled gluonia from $SU(\nd)$ SYM, the scale of their masses is $\sim\langle\lym^{(\rm br1)}\rangle\ll m_Q$, see \eqref{(6.1)}, the tension of the confining string is $\sigma^{1/2}\sim\langle\lym^{(\rm br1)}\rangle\ll\mu^{\,\rm pole}_{q,2}\ll\mu^{\,\rm pole}_{q,1}$. \\
4) There are $\no^2$ mions $M^1_1$ with masses $\mu^{\,\rm pole}(M^1_1)\ll m_Q$, see \eqref{(6.15)}.\\
5) There are $\nt^2$ mions $M^2_2$ with masses $\mu^{\,\rm pole}(M^2_2)\ll\langle\lym^{(\rm br1)}\rangle$, see \eqref{(6.16)}.\\
6) There are $2\no\no$ massless Nambu-Goldstone multiplets of $M^1_2$ and $M^2_1$.\\
7) All $N_F^2$ fions $\Phi^j_i$ have masses $\mu^{\rm pole}(\Phi)\sim\mph\gg\la$\,.\\

Comparing the mass spectra of the direct theory in section 6.1 and dual one in this section in these br1-vacua it is seen that they are parametrically different.

\section{Conclusions}

This paper finishes our study in \cite{ch4,ch5,ch6,ch7} of ${\cal N}=1$ SQCD-like theories (and their Seiberg's dual) with $N_F^2$ additional colorless but flavored scalar multiplets $\Phi^j_i$, with the large mass parameter $\mph\gg\la$. Therefore, we enumerate in these conclusions in a very short form only the main characteristic properties of such ${\cal N}=1$ SUSY theories which have been found out in these studies.\\

{\bf I)} The effective superpotential, see \eqref{(A.3)}, was proposed in section 3 of \cite{ch4} which accounts for all anomalies and depends only on quark bilinear combinations $\Pi^i_j=({\ov Q}_j Q^i)$. Together with \eqref{(A.2)} and Konishi anomalies \cite{Konishi}, see \eqref{(A.1)}, it allowed to calculate the values of gluino condensates and condensates of direct and dual quarks in numerous vacua of these theories, as well as multiplicities of these vacua, at different values of $N_F$ and in different regions of $\mph/\la$\,: a) at $1\leq N_F<2N_c$ in section 3 of \cite{ch4},\, b) at $N_F>2N_c$ in section 2 of \cite{ch6}. It appeared that the results in three different regions, $1<N_F<N_c\,,\, N_c<N_F<2N_c\,,\, N_F>2N_c$ are different, in general, and are not analytic continuation of each other. It was shown that the explicit dependence of various condensates on $(m_Q/\la)\ll 1$ and $(\mph/\la)\gg 1$, as well as multiplicities of various vacua, are qualitatively different in two regions\,: 1) $\la\ll\mph\ll\mo=\la(\la/m_Q)^{(2N_c-N_F)/N_c}$,\, 2) $\mph\gg\mo$, and the nontrivial evolutions of all condensates in different vacua at $\mph\lessgtr\mo$ were traced.\\

{\bf II)} A new nontrivial phenomenon was discovered in the direct theory in the strongly coupled regimes.
1) Due to the strong powerlike RG evolution at $\mu<\la$ with $N_c<N_F<3N_c$, the seemingly heavy and dynamically irrelevant at scales $\mu<\la$ colorless fields $\Phi^j_i$ with $\mph\gg\la$ can really become light and relevant, and there appear then two additional generations of light $\Phi$-particles with masses $\mu^{\rm pole}(\Phi)\ll\la$. This was described in detail in section 4 of \cite{ch4}, and in \cite{ch7,ch6,ch5} in different vacua of direct theories in strongly coupled or conformal regimes. 2) At $N_F>3N_c$ and $\mu>\la$ in the direct theory, for the same reasons, the seemingly heavy (i.e. with the large formal mass parameter $\mu_{\Phi}\gg\la$ in the superpotential) and dynamically irrelevant fields $\Phi^j_i$ can become very light and even massless, as described in this paper.\\

{\bf III)} The mass spectra in the direct theories with $SU(N_c)$ colors and in their Seiberg's dual ones with $SU(N_F-N_c)$ colors were calculated in \cite{ch5,ch7} in all different vacua at $1\leq N_F < 2N_c$ and $\mph\lessgtr\mo$, and in \cite{ch6} and this paper at $N_F>2N_c$ and $\mo\ll\la\ll\mph$.

At $1\leq N_F<N_c$ all quark masses and a part of gluon masses (except for those in remaining unbroken at scales $\mu>\lym$ SYM subgroup) in the direct $SU(N_c)$ theory are $\gg\la$, i.e. in the region where the theory is weakly coupled. Therefore, the calculations of mass spectra in this case are straightforward and there is no need for additional assumptions about the dynamics.

At $N_F>N_c$ either direct or dual theories are strongly coupled in most vacua. Therefore, at present, to calculate mass spectra in the strong coupling regime, one has to introduce some assumptions about the dynamics of such ${\cal N}=1$ theories. Calculations were performed in \cite{ch5,ch6,ch7} and in this paper in the framework of the dynamical scenario introduced in \cite{ch3} for the ordinary ${\cal N}=1$ SQCD (i.e. without additional fields $\Phi^j_i$, but with quarks of different masses). This scenario assumes that quarks in such ${\cal N}=1$ SQCD-like theories can be in two {\it standard} phases only\,: these are either the HQ (heavy quark) phase where they are not higgsed but confined, with $\langle Q\rangle=\langle{\ov Q}\rangle=0$, or the Higgs phase where they form nonzero coherent condensate with (at least some components of) $\langle Q^i_a\rangle=\langle{\ov Q}_i^a\rangle\neq 0$, breaking the color symmetry. The word {\it standard} implies here also that, unlike ${\cal N}=2$ SQCD, in such ${\cal N}=1$  theories without elementary colored adjoint scalars, no additional parametrically lighter solitons (e.g. magnetic monopoles or dyons) are formed at those scales where quarks decouple as heavy or are higgsed (see also the footnote \ref{(f2)}).

The calculations of mass spectra of SQCD-like theories \eqref{(1.1)},\eqref{(1.2)} were performed in the framework of this scenario in\,: \cite{ch5} for $3N_c/2<N_F<2N_c$, in \cite{ch6} for $2N_c<N_F<3N_c$, in \cite{ch7} for $N_c<N_F<3N_c/2$, and in this paper for $N_F>3N_c$. Recall that at $3N_c/2<N_F<3N_c$ both direct and dual theories considered in \cite{ch5,ch6} enter smoothly at $\mu<\la$ the strongly coupled (except for the parametrically narrow end regions) conformal regime with all particles of the direct and dual theories remaining effectively massless at $\mu<\la$. At $N_c<N_F<3N_c/2$ and $\mu\ll\la$ the direct theories are in the (very) strongly coupled regime with the gauge coupling $a(\mu\ll\la)\sim (\la/\mu)^{\,\nu_Q\,>\,0}\gg 1$, while dual theories are in the logarithmically weakly coupled regime with the dual gauge coupling ${\ov a}(\mu\ll\la)\sim 1/\log(\la/\mu)\ll 1$. Vice versa, at $N_F>3N_c$ the dual theories are logarithmically weakly coupled at $\mu\gg\la$ and (very) strongly coupled at $\mu\ll\la$, with the dual gauge coupling ${\ov a}(\mu\ll\la)\sim (\la/\mu)^{\,\nu_q\,>\,0}\gg 1$, while the direct theories are (very) strongly coupled at $\mu\gg\la$ and logarithmically weakly coupled at $\mu\ll\la$.

For these reasons, the mass spectra, in given theory and in a given vacuum, are parametrically different in different regions of $N_F$, as well as at $\mph\lessgtr\mo$. Besides, they are parametrically different in numerous different vacua. All these mass spectra were calculated and described in detail for both direct and dual theories in \cite{ch5,ch6,ch7} and in this paper for the whole region $N_F>N_c$.

In each given vacuum, comparison of mass spectra of direct theories and their Seiberg's dual ones showed that they are {\it  parametrically different}. In connection with this we note that, as it is seen from \cite{ch3} for the ordinary ${\cal N}=1$ SQCD and from \cite{ch5,ch6,ch7} for SQCD-like theories with additional $N_F^2$ colorless scalar multiplets $\Phi^j_i$, all results obtained within the used dynamical scenario from \cite{ch3} look self-consistent. In other words, no internal inconsistencies were encountered in all cases considered. Moreover, it is worth to remind that this dynamical scenario satisfies all those tests which were used as checks of the Seiberg hypothesis about the equivalence of the direct and dual theories. These parametrical differences of mass spectra of direct and dual theories show, in particular, that all these tests, although necessary, may well be insufficient.\\

{\bf IV)} In section 8 of \cite{ch6} were traced also close relations of these ${\cal N}=1$ SQCD-like theories with the ${\cal N}=2$ SQCD. Namely, with corresponding matching of parameters, all gluino and quark condensates in each vacuum, as well as multiplicities of various vacua, in these ${\cal N}=1$ SQCD-like theories with $N_F^2$ additional colorless scalars $\Phi^j_i$ with $\mph\gg\la$ are the same as in the broken ${\cal N}=2$ SQCD with the mass parameter $\mu_{\rm x}$ of adjoint colored scalars. Besides, the equality of mass spectra in br2 and S-vacua of softly broken ${\cal N}=2$ SQCD with $m\ll\mu_{\rm x}\ll\Lambda_2$ and in corresponding (with especially chosen parameters) Seiberg's dual variant of the ${\cal N}=1$ SQCD-like theory with $N_F^2$ additional colorless scalars $\Phi^j_i$ was shown. Also was emphasized that, in spite of the equality of these two mass spectra, these direct and Seiberg's dual theories have different t' Hooft triangles and so are not equivalent, in contrast with the claim in \cite{SY1} about their equivalence.

And finally, in this section 8 of \cite{ch6} the extensive criticism was presented of attempts made in \cite{SY2} to derive, similarly to \cite{APS}, the Seiberg duality of the ordinary ${\cal N}=1$ SQCD starting from ${\cal N}=2$ SQCD with $\mu_{\rm x}\ll\Lambda_2$ and evolving then to $\mu_{\rm x}\gg\Lambda_2$.

\appendix
\numberwithin{equation}{section}
\section{\hspace*{-2mm} Condensates and multiplicity of vacua at $N_F>2N_c$ \cite{ch6}}

\hspace*{5mm} For the reader convenience, we reproduce here first some useful formulae for the theories \eqref{(1.1)},\eqref{(1.2)} from section 3 in \cite{ch4} and from section 2 in \cite{ch6}.

The Konishi anomalies \cite{Konishi} for the $i$-th flavor look in the direct $\Phi$ - theory \eqref{(1.1)} as \cite{ch4}
\bbq
\langle\Phi_{i}\rangle\langle\frac{\partial {\cal W}_{\Phi}}{\partial \Phi_{i}}\rangle=0\,,\quad
\langle m_{Q,i}^{\rm tot}\rangle\langle {\ov Q}_i Q^i\rangle=\langle S\rangle\,,\quad \langle m_{Q,\,i}^{\rm tot}\rangle=m_Q-\langle\Phi_{i}\rangle\,,\quad \langle\Phi^j_{i}\rangle=\delta^j_i \langle\Phi_{i}\rangle,
\eeq
\bq
\langle\Phi_j^i\rangle=\frac{1}{\mph}\Biggl ( \langle{\ov Q}_j Q^i \rangle-\delta^i_j\frac{1}{N_c}{\rm Tr}\,\langle\qq\rangle\Biggr )\,,\quad \langle{\ov Q}_j Q^i \rangle\equiv\sum_{a=1}^{N_c}\langle{\ov Q}^{\,a}_j Q^i_{a}\rangle =\delta^i_j\langle{\ov Q}_i Q^i \rangle\,,\quad {\it i}=1\, ...\, N_F\,, \label{(A.1)}
\eq
and $\langle m_{Q,i}^{\rm tot}\rangle$ is the value of the quark running mass at the scale $\mu=\la$.

At all scales until the field $\Phi$ remains too heavy and non-dynamical, i.e. until its perturbative running mass $\mu_{\Phi}^{\rm pert}(\mu)>\mu$, it can be integrated out and the superpotential takes the form
\bq
{\cal W}_Q=m_Q{\rm Tr}({\ov Q} Q)-\frac{1}{2\mph}\Biggl (\,\sum_{i,j}\,({\ov Q}_j Q^i)({\ov Q}_i Q^j)-\frac{1}{N_c}\Bigl({\rm Tr}\,{\ov Q} Q \Bigr)^2 \Biggr ). \label{(A.2)}
\eq

The values of the quark condensates for the $i$-th flavor, $\langle{\ov Q}_i Q^i\rangle$, in various vacua can be obtained from the effective superpotential \cite{ch4} which accounts for all anomalies and depends only on quark bilinear combinations $\Pi^i_j=({\ov Q}_j Q^i)$, see \eqref{(A.1)} for $({\ov Q}_j Q^i)$ and \eqref{(A.2)} for ${\cal W}_Q$,
\bq
{\cal W}_{\rm eff}(\Pi)=-\nd S+W_Q\,,\quad S=\Bigl (\frac{\det{\ov Q} Q}{\la^{\bo}}\Bigr )^{1/\nd}\,,\quad \bo=3N_c-N_F\,,\quad \nd=N_F-N_c\,. \label{(A.3)}
\eq

For the vacua with the spontaneously broken flavor symmetry, $U(N_F)\ra U(n_1)\times U(n_2)$, the most convenient way to find the quark condensates is to use \cite{ch4}
\bbq
\langle \Qo+\Qt-\frac{1}{N_c}{\rm Tr}\, ({\ov Q} Q)\,\rangle_{\rm br}=m_Q\mph,\quad
\langle S\rangle_{\rm br}=\Bigl (\frac{\det \langle{\ov Q} Q\rangle_{\rm br}}{\la^{\bo}}\Bigr )^{1/\nd}=\frac{\langle\Qo \rangle_{\rm br}\langle\Qt\rangle_{\rm br}}{\mph},
\eeq
\bq
\det \langle{\ov Q} Q\rangle_{\rm br}=\langle\Qo\rangle^{{\rm n}_1}_{\rm br}\,\langle\Qt\rangle^{{\rm n}_2}_{\rm br}\,,\quad \Qo\equiv\sum_{a=1}^{N_c}{\ov Q}^{\,a}_1 Q^1_{a}\,,\quad \Qt\equiv\sum_{a=1}^{N_c}{\ov Q}^{\,a}_2 Q^2_{a}\,, \label{(A.4)}
\eq
\bbq
\langle m^{\rm tot}_{Q,1}\rangle_{\rm br}=m_Q-\langle\Phi_1\rangle_{\rm br}=\frac{\langle\Qt\rangle_{\rm br}}{\mph},\quad \langle m^{\rm tot}_{Q,2}\rangle_{\rm br}=m_Q-\langle\Phi_2\rangle_{\rm br}=\frac{\langle\Qo\rangle_{\rm br}}{\mph}\,.
\eeq

The Konishi anomalies for the $i$-th flavor look in the dual $d\Phi$ - theory \eqref{(1.2)} as
\bq
\langle M_i\rangle \langle {\ov q}^{\,i} q_i\rangle=\la\langle S\rangle\,,\quad \frac{\langle{\ov q}^{\,i} q_i\rangle}{\la}=\langle m_{Q,i}^{\rm tot}\rangle=m_Q-\frac{1}{\mph}\,\langle M_i-\frac{1}{N_c}{\rm Tr}\,M \rangle\,,\quad {\it i}=1\, ...\, N_F\,. \label{(A.5)}
\eq

In vacua with the broken flavor symmetry these can be rewritten as (remind that $\langle M^i_j\rangle=
\delta^i_j\langle M_i\rangle,\\ \langle M_1\rangle=\langle\Qo\rangle,\, \langle M_2\rangle=\langle\Qt\rangle$)
\bbq
\langle M_1+M_2-\frac{1}{N_c}{\rm Tr}\, M\rangle_{\rm br}=m_Q\mph,\quad\langle S\rangle_{\rm br}=\Bigl (\frac{\det \langle M\rangle_{\rm br}}
{\la^{\bo}}\Bigr )^{1/\nd}=\frac{1}{\mph}\langle M_1\rangle_{\rm br}\langle M_2\rangle_{\rm br}\,,
\eeq
\bq
\frac{\langle\qo\rangle_{\rm br}}{\la}=\frac{\langle S\rangle_{\rm br}}{\langle M_{1}\rangle_{\rm br}}=\frac{\langle M_{2}\rangle_{\rm br}}{\mph}=\langle m^{\rm tot}_{Q,1}\rangle_{\rm br},\quad
\frac{\langle \qt\rangle_{\rm br}}{\la}=\frac{\langle S\rangle_{\rm br}}{\langle M_{2}\rangle_{\rm br}}=\frac{\langle M_{1}\rangle_{\rm br}}{\mph}=\langle m^{\rm tot}_{Q,1}\rangle_{\rm br}\,, \label{(A.6)}
\eq
\bbq
\frac{\langle\qt\rangle_{\rm br}}{\langle\qo\rangle_{\rm br}}=\frac{\langle\Qo\rangle_{\rm br}}{\langle\Qt \rangle_{\rm br}}\,,\quad \qo\equiv\sum_{b=1}^{\nd}{\ov q}_{b}^{\,1} q^{b}_1\,,\quad \qt\equiv\sum_{b=1}^{\nd} {\ov q}_{b}^{\,2} q^b_2\,,\quad \nd\equiv N_F-N_c\,.
\eeq

\subsection{Vacua with the unbroken flavor symmetry}

One obtains from \eqref{(A.3)} at $\mph\lessgtr \mo$ and with $\langle{\ov Q}_j Q^i\rangle=\delta^i_j\langle{\ov Q} Q\rangle\,,\,\, \langle{\ov Q} Q\rangle=\sum_{a=1}^{N_c}{\ov Q}^{\,a}_1 Q^1_{a}\,$. -

{\bf a)} There are only $\nd=(N_F-N_c)$ nearly classical S-vacua (S=small) at $\mph\ll\mo$ with
\bq
\langle\qq\rangle_S\equiv\langle\qq(\mu=\la)\rangle_S\simeq -\frac{N_c}{\nd}\, m_Q\mph\,,\quad \mo=\la\Bigl (\frac{m_Q}{\la}\Bigr )^{\frac{N_F-2N_c}{N_c}}\ll\la\,, \label{(A.7)}
\eq
\bbq
\langle\lym^{(S)}\rangle^3\equiv\langle S\rangle_S=\Bigl (\frac{\det\langle\qq\rangle_S}{\la^{\rm \bo}}\Bigr )^{1/\nd}\sim\la^3\Bigl (\frac{m\mph}{\la^2}\Bigr )^{N_F/\nd}\,.
\eeq

{\bf b)} There are $(N_F-2N_c)$ quantum L-vacua (L=large) at $\mph\gg\mo$ with
\bq
\langle\qq\rangle_L\equiv\langle\qq(\mu=\la)\rangle_L\sim \la^2\Biggl (\frac{\mph}{\la}\Biggr )^{\frac{\nd}{N_F-2N_c}}\,, \label{(A.8)}
\eq
\bbq
\langle\lym^{(L)}\rangle^3\equiv\langle S\rangle_L=\Bigl (\frac{\det\langle\qq\rangle_L}{\la^{\rm \bo}}\Bigr )^{1/\nd}\sim\la^3\Bigl (\frac{\mph}{\la}\Bigr )^{\frac{N_F}{N_F-2N_c}}\,.
\eeq

{\bf c)} There are $N_c$ quantum QCD  vacua at $\mph\gg\mo$ with
\bq
\langle\qq\rangle_{QCD}=\langle\qq(\mu=\la)\rangle_{QCD}\simeq\frac{\langle S\rangle_{\rm QCD}\equiv \langle\lym^{(\rm QCD)}\rangle^3}{m_Q}\simeq\frac{1}{m_Q}\Bigl (\la^{\bo}m_Q^{N_F}\Bigr)^{1/N_c}\,. \label{(A.9)}
\eq

The total number of vacua with the unbroken flavor symmetry is
\bq
N^{\rm tot}_{\rm unbr}=(N_F-2N_c)+N_c=\nd\,. \label{(A.10)}
\eq

\subsection{Vacua with the broken flavor symmetry \\
\hspace*{1cm} $U(N_F)\ra U(n_1)\times U(n_2),\quad 1\leq n_1\leq [N_F/2]$}

{\bf a)} There are $(n_1-N_c){\ov C}^{\,\rm n_1}_{N_F}$ br1-vacua
\footnote {\,
${\ov C}^{\, n_1}_{N_F}$ differ from the standard $C^{\, n_1}_{N_F}=N_F!/[n_1!\,n_2!]$ only for ${\ov C}^{\,n_1={\rm k}}_{N_F=2{\rm k}}=C^{\,n_1={\rm k}}_{N_F=2{\rm k}}/2$.

Besides, by convention, we ignore the continuous multiplicity of vacua due to the spontaneous flavor symmetry breaking. Another way, one can separate slightly all quark masses, so that all Nambu-Goldstone particles will acquire small masses $O(\delta m_Q)\ll {\ov m}_Q$.
}
(br=breaking, with the dominant $\langle\Qo\rangle$)
at $N_c<n_1\leq [N_F/2]$ and $\mph\ll\mo$ with
\bq
\langle\Qo\rangle_{\rm br1}\simeq \frac{N_c}{N_c-n_1} m_Q\mph,\quad  \langle\Qt \rangle_{\rm br1}\sim \la^2\Bigl (\frac{\mph}{\la}\Bigr )^{\frac{n_1}{n_1-N_c}}\Bigl (\frac{\la}{m_Q}\Bigr )^{\frac{n_2-N_c}{n_1-N_c}}\,, \label{(A.11)}
\eq
\bbq
\langle S\rangle_{\rm br1}=\frac{\langle\Qo\rangle_{\rm br1}\langle\Qt\rangle_{\rm br1}}{\mph}\sim
\la^3\Bigl (\frac{\mph}{\la}\Bigr )^{\frac{n_1}{n_1-N_c}}\Bigl (\frac{\la}{m_Q}\Bigr )^{\frac{n_2-n_1}{n_1-N_c}}\,,\quad\frac{\langle\Qt\rangle_{\rm br1}}{\langle\Qo\rangle_{\rm br1}}\sim \Bigl (\frac{\mph}{\mo}\Bigr )^{\frac{N_c}{n_1-N_c}}\ll 1\,.
\eeq

{\bf b)} There are $(n_2-N_c){\ov C}^{\,\rm n_1}_{N_F}$ br2 -vacua at all values $1\leq n_1\leq [N_F/2]$ and $\mph\ll\mo$ with
\bq
\langle\Qt\rangle_{\rm br2}\simeq \frac{N_c}{N_c-n_2} m_Q\mph\,,\quad \langle\Qo\rangle_{\rm br2}\sim \la^2\Bigl (\frac{\mph}{\la}\Bigr )^{\frac{n_2}{n_2-N_c}}\Bigl (\frac{m_Q}{\la}\Bigr )^{\frac{N_c-n_1}{n_2-N_c}}\,,\label{(A.12)}
\eq
\bbq
\langle\lym^{(\rm br2)}\rangle^3\equiv\langle S\rangle_{\rm br2}\sim\la^3\Bigl (\frac{\mph}{\la}\Bigr )^{\frac{n_2}{n_2-N_c}}\Bigl (\frac{m_Q}{\la}\Bigr )^{\frac{n_2-n_1}{n_2-N_c}}\,,\quad
\frac{\langle\Qo\rangle_{\rm br2}}{\langle\Qt\rangle_{\rm br2}}\sim \Bigl (\frac{\mph}{\mo}\Bigr )^{\frac{N_c}{n_2-N_c}}\ll 1\,.
\eeq

On the whole, there are ($\,\theta(z)$ is the step function)
\bq
N^{\rm tot}_{\rm br}(n_1)=\Bigl [ (n_2-N_c)+\theta(n_1-N_c)(n_1-N_c)\Bigr ]{\ov C}^{\,\rm n_1}_{N_F},
\,\, N^{\rm tot}_{\rm br}=\sum_{n_1=1}^{[N_F/2]}N^{\rm tot}_{\rm br}(n_1)\,,\,\, \no+\nt=N_F \label{(A.13)}
\eq
vacua with the broken flavor symmetry at $\mph\ll\mo\,$.

\vspace{2mm}

{\bf c)} There are $(N_c-n_1)C^{\rm\,n_1}_{N_F}$ br1 -vacua at $1\le n_1<N_c$ and $\mph\gg\mo$ with
\bq
\langle\Qo\rangle_{\rm br1}\simeq \frac{N_c}{N_c-n_1} m_Q\mph,\quad\langle\Qt\rangle_{\rm br1}\sim \la^2\Bigl (\frac{\la}
{\mph}\Bigr )^{\frac{n_1}{N_c-n_1}}\Bigl (\frac{m_Q}{\la}\Bigr )^{\frac{n_2-N_c}{N_c-n_1}}\,, \label{(A.14)}
\eq
\bbq
\langle\lym^{(\rm br1)}\rangle^3\equiv\langle S\rangle_{\rm br1}\sim\la^3\Bigl (\frac{\la}{\mph}\Bigr )^{\frac{n_1}{N_c-n_1}}\Bigl (\frac{m_Q}{\la}\Bigr )^{\frac{n_2-n_1}{N_c-n_1}}\,,\quad
\quad \frac{\langle\Qt\rangle_{\rm br1}}{\langle\Qo\rangle_{\rm br1}}\sim \Bigl (\frac{\mo}{\mph}\Bigr )^{\frac{N_c}{N_c-n_1}}\ll 1\,.
\eeq

{\bf d)} There are $(N_F-2N_c){\ov C}^{\rm\,n_1}_{N_F}$ Lt (Lt=L-type) vacua  at $n_1\neq N_c$ and $\mph\gg\mo$ with
\bq
(1-\frac{n_1}{N_c})\langle\Qo\rangle_{\rm Lt}\simeq -(1-\frac{n_2}{N_c})\langle\Qt\rangle_{\rm Lt}\sim \la^2\Biggl (\frac{\mph}{\la}\Biggr )^{\frac{\nd}{N_F-2N_c}}, \label{(A.15)}
\eq
i.e. as in the L-vacua in (A.8) but $\langle\Qo\rangle_{\rm Lt}\neq\langle\Qt\rangle_{\rm Lt}$ here.

{\bf e)} There are $(N_F-2N_c) C^{\rm\,n_1}_{N_F}$ special vacua at $n_1=N_c$ and $\mph\gg\mo$ with
\bbq
\langle\Qt\rangle_{\rm spec}=\frac{N_c}{2N_c-N_F} m_Q\mph,\,\,\langle\Qo\rangle_{\rm spec}=\la^2\Bigl (\frac{\mph}{\la}\Bigr )^{\frac{\nd}{N_F-2N_c}}\,,
\eeq
\bq
\langle S\rangle_{\rm spec}\sim m_Q\la^2\Bigl (\frac{\mph}{\la}\Bigr )^{\frac{\nd}{N_F-2N_c}}\,,\quad \frac{\langle\Qt\rangle_{\rm spec}}{\langle\Qo\rangle_{\rm spec}}\sim \Bigl (\frac{\mo}{\mph}\Bigr )^{\frac{N_c}{N_F-2N_c}}\ll 1\,.\label{(A.16)}
\eq

As one can see from the above, similarly to \cite{ch4} with $N_c<N_F<2N_c$, all quark condensates become parametrically the same at $\mph\sim\mo$, see \eqref{(A.7)}. Clearly, just this region $\mph\sim\mo$ and not $\mph\sim \la$ is very special and most of the quark condensates change their parametric behavior and hierarchies at $\mph\lessgtr\mo$. For example, the br1 - vacua with $1\le n_1<N_c\,,\,\,\langle\Qo \rangle\sim m_Q\mph\gg\langle\Qt\rangle$ at $\mph\gg\mo$ evolve into br2 - vacua with $\langle\Qt\rangle\sim m_Q\mph\gg\langle\Qo\rangle$ at $\mph\ll\mo$, while the br1 - vacua with $n_1>N_c\,,\,\,\langle\Qo\rangle\sim m_Q\mph\gg\langle\Qt\rangle$ at $\mph\ll\mo$ evolve into the L-type  vacua with $\langle\Qo\rangle
\sim\langle\Qt\rangle\sim\la^2\, (\mph/\la)^{\nd/(N_F-2N_c)}$ at $\mph\gg\mo$, etc.

The exception is the special vacua with $n_1=N_c\,,\, n_2=\nd$\,. In these, the parametric behavior $\langle\Qt\rangle\sim m_Q\mph, \,\langle\Qo\rangle\sim \la^2\,(\mph/\la)^{\nd/(N_F-2N_c)}$ remains the same but the hierarchy is reversed at $\mph\lessgtr\mo\, :\, \langle\Qo\rangle/\langle\Qt\rangle\sim (\mph/\mo)^{N_c/(N_F-2N_c)}$.\\

On the whole, there are
\bq
N^{\rm tot}_{\rm br}(n_1)=\Bigl [ (N_F-2N_c)+\theta(N_c-n_1)(N_c-n_1)\Bigr ]{\ov C}^{\,\rm n_1}_{N_F},
\quad N^{\rm tot}_{\rm br}=\sum_{n_1=1}^{[N_F/2]}N^{\rm tot}_{\rm br}(n_1)
\eq
vacua with the broken flavor symmetry at $\mph\gg\mo$. The total number of vacua is the same at $\mph\lessgtr\mo$\,, as it should be.

We point out finally that the multiplicities of vacua at $N_F>2N_c$ are not the analytic continuations of those at $N_c<N_F<2N_c$\,, see section 3 in \cite{ch4}.

\end{document}